%% file: main_final.tex
\documentclass[journal]{IEEEtran}
\usepackage{amsmath,amssymb,amscd,latexsym,dsfont}
\usepackage{comment}
\usepackage{color}
\usepackage{subfigure}
\usepackage{enumerate}
\usepackage{enumitem}
\usepackage{stfloats}
\usepackage{psfrag} 
\usepackage{setspace}
\usepackage{cite}
\usepackage{balance}
\usepackage{enumitem}
\usepackage{tikz}
\usepackage{pgfplots,amsfonts}
\usepackage{pgfplotstable}
\usetikzlibrary{shapes}
\usetikzlibrary{spy}
\usetikzlibrary{fit}
\usetikzlibrary{shapes.multipart}
\usetikzlibrary{positioning}
\input{myFigures.tex}

\input{myStyles.tex}
\input{myMacros.tex}
\input{myPackages.tex}
\pgfplotsset{compat=1.14}
\usepackage{wrapfig}
\usepackage[acronym,nomain]{glossaries}
\input{acronyms.tex}
\usepackage{arydshln}

\usepgfplotslibrary{fillbetween}
\usepackage{tikz-network} % provides graph / network utilities (Edge, Node, etc)

\usetikzlibrary{arrows.meta} % for nice arrows
\usetikzlibrary{calc} % to do some computations on the coordinates

\usepackage{colortbl} % to color rows or columns of matrices

\usetikzlibrary{backgrounds} % to explicitely draw in the background layer

\input{colors.tex}

\usetikzlibrary{arrows}
\usetikzlibrary{3d}

\usepackage{enumerate}

\title{Geometrically-Shaped Multi-Dimensional Modulation Formats in Coherent Optical Transmission Systems
}

\author{Bin~Chen,~\IEEEmembership{Senior Member,~IEEE}, Yi Lei,~\IEEEmembership{Member,~IEEE}, Gabriele Liga,~\IEEEmembership{Member,~IEEE}, \\ Zhiwei Liang,  Wei Ling, Xuwei Xue,~\IEEEmembership{Member,~IEEE}, and Alex~Alvarado,~\IEEEmembership{Senior Member,~IEEE}

\thanks{The work of B. Chen and Y. Lei are   supported by the National Natural Science Foundation of China  (No. 62171175, 62001151 and 62101065), by the Fundamental Research Funds for the Central Universities (No. JZ2022HGTB0262).
The work of G. Liga is funded by the EuroTechPostdoc programme and  the European Research Council  under the European Union's Horizon 2020 research and innovation programme (No. 754462).
The work of A. Alvarado is supported by the Netherlands Organisation for Scientific Research (NWO) via the VIDI Grant ICONIC (project number 15685). 
Parts of this paper have been presented at the \textit{Optical Fiber Communication Conference and Exhibition (OFC)}, San Diego, California, USA, 2022 \cite{ChenOFC2022}.}
\thanks{B. Chen, Y. Lei, Z. Liang and W. Ling   are with the School of Computer Science and Information Engineering, Hefei University of Technology, Hefei, China (e-mails:~\{bin.chen,leiyi\}@hfut.edu.cn). %\{2021111029,2020171108\}@mail.hfut.edu.cn).
}
\thanks{G. Liga and A. Alvarado are with the Department of Electrical Engineering, Eindhoven University of Technology, Eindhoven, The Netherlands (e-mails:~\{g.liga,a.alvarado\}@tue.nl).}
\thanks{X. Xue is with State Key Laboratory of Information Photonics and Optical Communications, Beijing University of Posts and Telecommunications, Beijing, China (e-mail:~x.xue@bupt.edu.cn).}
}

\begin{document}
%\bstctlcite{IEEEexample:BSTcontrol}

\maketitle

\begin{abstract}
Shaping  modulation formats in multi-dimensional (MD) space is an effective approach to harvest spectral efficiency gains in both the additive white Gaussian noise  (AWGN) channel and the optical fiber channel. 
In the first part of this paper, existing MD geometrically-shaped modulations for fiber optical communications are reviewed.
It is shown that large gains can be obtained by exploiting correlation  in the dimensions or/and by increasing the  cardinality of the  modulation format. 
Practical limitations and challenges are also discussed together with efficient solutions. 
In the second part, we  extend the recently proposed four-dimensional (4D) modulation format family based on the constraint of orthant-symmetry to   high spectrum efficiencies  up to 10~bit/4D-sym by maximizing generalized mutual information for AWGN channel. Reach increases of up to 25\% for a multi-span  optical fiber transmission system are reported.
Lastly, with the help of a recently introduced nonlinear interference (NLI) model, an optimization for designing nonlinear-tolerant 4D modulation formats is introduced for a single-span optical fiber system. Simulation results show that the proposed NLI model-based 4D modulation format could  increase the effective SNRs by 0.25~dB  with respect to the AWGN channel-optimal 4D  modulation format.
\end{abstract}


\section{Introduction}
{Currently, one of the main challenges in optical fiber transmission systems is to ensure that the available capacity within the deployed systems is efficiently utilized. A key technique to achieve this is to combine forward error correction (FEC) with nonbinary modulation formats, which is known as coded modulation (CM) \cite{UngerboeckTIT1982}.   
For a robust transmission system, a trade-off between data rate, noise tolerance, demapping-decoding computational complexity has to be made when optimizing constellations for a given transmission distance.
400G (and emerging 800G) optical CM transceivers use polarization-multiplexed (PM) two-dimensional (2D) constellations such as 16-ary quadrature amplitude modulation (QAM) and 64QAM, however, other alternatives exist.}

%% signal shaping 
Signal shaping has recently been widely investigated in optical fiber communications to improve spectral efficiency (SE), and is currently implemented in commercial products via probabilistic shaping (PS) \cite{NokiaPSE-V} and geometric shaping (GS) \cite{FujitsuT600}. 
Both PS and GS are often used to mimic a Gaussian distribution on the symbols by either changing the probability or the position of each constellation point compared to conventional rectangular structures (i.e., QAM formats). 
{Both techniques have distinct advantages and disadvantages.
PS has been shown to offer a near-optimal linear shaping gain and a greater flexibility  than GS in general \cite[Sec.~4.2]{Szczecinski2015BICM}, \cite{SteinerSCC2017}. In addition, by applying  PS to a standard QAM constellation, the optimum  Gray-labeling can easily employed. 
However, PS requires an external distribution matcher (DM) with an efficient  implementation  and this drawback of PS can limit its shaping gain. 
GS  only requires straightforward modifications of the mapper and demapper, which can  be easily coupled with FEC and designed for  different impairments  (e.g., fibre nonlinearity \cite{SillekensECOC2018}  and laser phase noise \cite{HubertJLT2021}).
However, the  irregular constellation points of GS  requires a finer digital-to-analog converter (DAC) and analog-to-digital converter (ADC) precision, and also  increases the computational complexity of the demapper.
Furthermore, in practical optical fiber system, both PS and GS suffer from rate loss due to the
nonlinear effects of the fiber and practical implementation penalty.
Research is therefore now dedicated to find improved PS, GS and  hybrid PS/GS architectures.
}

The standard approach in optical fiber systems consists in encoding data independently over each polarization channel using the same 2D modulation format, leading to polarization-multiplexed 2D (PM-2D) formats. 
For 2D constellations, GS optimizations were already proposed in the 70's in the communication theory literature \cite{WeltiTIT1974,FoschiniTCOM1974,ZetterbergTCOM1977}. 
These optimizations were more recently revisited in the context of optical fiber transmission in \cite{LotzJLT2013,ZhangECOC2017,BinECOC2018,BinICTON2018}.
A smaller but still significant amount of work was also performed on  constellation design  in optical fiber channel with nonlinearities \cite{RasmusECOC2018,ShenLIECOC2018,SillekensECOC2018,SchaedlerOFC020,Rasmus2019endtoend,AmirhoseinJLT2022} and laser phase noise \cite{HubertJLT2021,jovanovic2021ECOC,RodeOFC2022}.
By using 2D GS formats, a record transmission throughput of  178.08~Tbit/s over 40~km and  a record capacity transmission of 74~Tbit/s over 6300~km were reported in \cite{GaldinoPTL2020} and \cite{MariaJLT2020}, respectively.
{These 2D GS formats are listed in {the first part} of  Table~\ref{tab:MD_GS}, which is a  broad, but nonexhaustive  overview of existing GS works for optical communication systems}.

Even though 2D geometric shaping has been shown to increase the achievable information rates (AIRs) in optical fiber systems,  
these modulations with a finite number of constellation points perform suboptimally in terms of AIR when compared to probabilistically-shaped QAM  in the AWGN channel \cite[Sec.~4.2]{Szczecinski2015BICM}, \cite{SteinerSCC2017}. %
For both the AWGN and the optical fiber channel, higher shaping gains or/and coding gains are to be expected from modulation formats with a higher number of dimensions 
\cite{ForneyJSAC1984,LotfollahSPM2014}. 
Thus, the design of MD modulation formats  has been considered as  an effective approach to harvest performance gains in optical communications. {These formats include for example 4D, 8D, 12D, 16D, etc., as shown in Table~\ref{tab:MD_GS}.}

\begin{table*}[!tb]
\vspace{-0.5em}
		\centering
		\caption{Published gains obtained by Geometric shaping in optical fiber communication systems: 2D $\rightarrow$ 4D $\rightarrow$ 8D $\rightarrow$ MD.\protect\footnotemark}
		\small
		\vspace{-0.2em}
		\input{table/MD_GS_review}
		\label{tab:MD_GS}
		\vspace{-1.5em}
\end{table*}
\footnotetext{All the reported results in Table~\ref{tab:MD_GS} are compared with the baselines at the same SE and under the same FEC overhead assumption,  except PS-QPSK in \cite{Karlsson:09,AgrellJLT2009},  8PolSK–QPSK in \cite{Chagnon:13},  4D-512-Hurwitz in \cite{FreyJLT2020}, and the 8D formats in \cite{Bendimerad:18}, which however do make a fair comparison by modifying symbol rate or FEC overhead.}

\begin{figure}[!tb]
    \centering
{\includegraphics{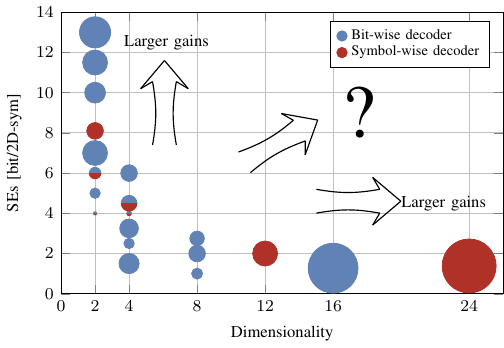}}
   \vspace{-1em}
   \caption{Illustration of approximate gains of MD geometrically-shaped modulation formats vs. SEs  dimensionality. %Red- and blue-colored markers indicate the target performance metrics for symbol-wise decoder and bit-wise  decoder, respectively. 
   Radii of the markers are proportional to the corresponding SNR gains in Table~\ref{tab:MD_GS} with respect to conventional QAM.
   }
    \label{fig:MD_gains}
       \vspace{-1.7em}
\end{figure}

Most of the so-called MD modulation formats in the early literature are in fact only optimized in each complex dimension independently. 
In this sense, they are not \textit{true} MD modulation formats. This is the case of  PM-2D formats, 
where two identical constellations are transmitted  independently over two orthogonal polarizations. 
A true MD format is not in general generated as Cartesian products of a component 1D or 2D constellation. %\cite{Agrell2014_JLT,Karlsson:09}.
One way to generate MD modulation formats with dependency between dimensions is applying Ungerboeck's  set-partitioning (SP) scheme \cite{UngerboeckTIT1982}, which maps binary bits onto multiple consecutive 2D symbols in a selected subset  to generate a MD symbol. 
For example, in four-dimensional (4D) space,  
set-partitioning PM-16QAM has been investigated to  achieve fine granularity  by mapping 5--7 bits onto two consecutive QAM symbols. These formats are known  as  32-ary SP 16QAM (32SP-QAM) \cite{MullerECOC2013,SunECOC2013}, 64-ary SP 16QAM (64SP-QAM) \cite{NakamuraECOC2015} and 128-ary SP 16QAM (128SP-QAM) \cite{ErikssonOE13,KashiECOC2015,KaihuiOFT2018}.  
4D set-partitioned  formats based on the  $\boldsymbol{D}_4$ lattice have also been investigated by using multilevel coding  and multi-stage decoding  \cite{FreyJLT2020}.

One of the main advantages of performing shaping in MD space is that it can also  be used to mitigate the nonlinear effects in the optical fiber channel.
This insight motivates the search for modulation formats in a higher dimensional space that are tolerant to linear noise and/or nonlinear interference (NLI)  \cite{Dar14_ISIT}.
Intuitively,  waveforms with less intensity variation generate lower NLI. Therefore, 4D constant modulus  constellations were first proposed to achieve  better nonlinearity tolerance performance compared to multi-modulus ones 
\cite{Karlsson:09,AgrellJLT2009,Chagnon:13,ReimerOFC2016,Kojima2017JLT,BinChenJLT2019,ReneECOC2020}. For example, polarization switched-QPSK (PS-QPSK) \cite{Karlsson:09,AgrellJLT2009} maps 3 bits onto two consecutive QPSK symbols,  while 4D2A8PSK \cite{Kojima2017JLT} maps 5--7 bits onto two consecutive 8PSK symbols in both X- and Y-polarization.  4D-64PRS \cite{BinChenJLT2019} is obtained by maximizing generalized
mutual information (GMI) under the constraint of 4D constant modulus and was transmitted over 11,700~km in a laboratory experiment with standard single-mode fiber (SSMF) \cite{SjoerdOECC2019}. 
{More generally, an MD modulation format can ``shape out" the partial NLI   at the expense of losing  some degrees of freedom \cite{Dar14_ISIT}.} Furthermore, increasing the dimensionality provides flexibility to optimize the constellation power efficiency  and introduces specific correlations between different dimensions that can further  reduce the generation of nonlinear effects during signal propagation.

MD modulation formats in  8D \cite{Shiner:14}, 12D \cite{ReneOFC2020}, 16D \cite{RademacherECOC2015}	 and 24D  \cite{millar2013SPPcom,Millar:14,millar2014OFC} have been proposed by adding proper constraints in the  optimization to mitigate modulation-dependent NLI interaction 
and to  extend the transmission reach. 
For example, the eight dimensions were obtained in  \cite{ErikssonECOC2013}  by combining two polarizations with two frequencies  and in \cite{Shiner:14,KoikeAkinoECOC2013} by using two consecutive time slots.
In \cite{Bendimerad:18,El-RahmanJLT2018,BinChenPTL2019,SjoerdECOC2019}, 8D modulation formats in two consecutive time slots were designed to further mitigate fiber nonlinear impairments using the polarization-balancing  or polarization-alternating  concept.
24D modulation formats in optical systems were also demonstrated based on  sphere cutting of lattices and block codes (as inner codes)  in \cite{millar2013SPPcom,Millar:14} and 15\% reach increase was shown  over BPSK in \cite{millar2014OFC}. 
{In \cite{ReneOFC2020}, a 12D modulation format   was demonstrated across three linearly coupled spatial modes of a multicore fiber  showing improved performance over PM-QPSK.}
To pack the constellation points more efficient,  ultra-large Voronoi constellations  with up to  $10^{28}$ MD points and larger SE were recently demonstrated in \cite{MiraniJLT2021} to show  significant bit error rate (BER) and symbol error rate (SER) gains over QAM.

The approximated gains in terms of SNR for  MD geometrically-shaped  modulation  in Table~\ref{tab:MD_GS} are shown   in Fig.~\ref{fig:MD_gains} for different dimensionalities and SEs. Fig.~\ref{fig:MD_gains} shows that (\romannumeral1)  large constellation sizes can potentially achieve large gains, 
(\romannumeral2)  high dimensionalities can potentially achieve large gains, 
(\romannumeral3)   research has mainly focused on the  optimization of constellations either in a low-dimensional space (with high SE), or on  high-dimensional constellations (but with relatively low SE). 
As shown in Fig.~\ref{fig:MD_gains}, the current challenge is the full optimization of the constellation in a high-dimensional space with  higher SEs to achieve large gains.

 The contributions of this paper are three. 
 First, {we provide a review of existing multidimensional geometric shaping (MD-GS) works to provide more insights, e.g., limitations and challenges, into   the employed MD-GS modulation formats in optical fiber systems.}
{In particular, we also discuss methods for enabling {efficient} MD-GS performance enhancement and complexity reduction to achieve both linear shaping gains and nonlinearity tolerance.} 
Secondly, in this paper we focus on designing 4D modulation formats for  soft-decision (SD) FEC with 20\%-25\% overhead in  bit-interleaved coded modulation (BICM) systems by maximizing the GMI and thus, increasing transmission reach.  
Simulation comparisons for a set of AWGN-optimized 4D formats with SEs of $5$--$10$~bits/4D-sym, which outperform PM-$M$QAM and the previously known 4D formats, are presented in a multi-span optical fiber communication system. 
Finally, to  design nonlinear-tolerant modulation formats in optical fiber systems, an optimization of dual-polarization (DP) modulation based on the  4D NLI model  \cite{GabrieleEntropy2020} 
is performed to mitigate NLI in a single-span transmission scenario.

The remainder of this paper is structured as follows.  In Sec.~\ref{sec:model}, we present the system model adopted for our constellation design and introduce the  performance metrics for the MD coded modulation system under consideration. In Sec.~\ref{sec:optimization}, we discuss  general aspects of MD geometric shaping. 
In  Sec.~\ref{sec:solutions}, we present the  methods for enabling efficient multi-dimensional modulation optimization and its implementation.
The simulation results of optical fiber transmission for multi-span system and single-span system  are further presented in Sec.~\ref{sec:multi_span} and Sec.~\ref{sec:single_span}, respectively.
 Finally, the paper is concluded in Sec.~\ref{sec:con}.

\textit{Notations}: {Bold symbols denote either  row vectors (lower case) or  random vectors (upper case).}
The elements of a vector $\bx$ are denoted by $x_i$,  and the element  at row $i$,
column $j$  of a matrix $\underline{\boldsymbol{x}}$ are denoted by $x_{i,j}$. Expectation denoted by $\mathbb{E}[ \cdot ]$ and  the squared Euclidean of $N/2$-dimensional complex vector is defined as $||\bX||^2=|X_1|^2+|X_2|^2+...+|X_N|^2$. 
The set is  denoted by $\mathcal{A}$ and the Cartesian product between sets is denoted by $\mathcal{A}\times\mathcal{A}$.

\begin{figure*}[!tb]
\vspace{-0.5em}
  \centering
%  \scalebox{0.9}{\input{./tikz/SystemModel.tikz}}
 \scalebox{0.9}{{\includegraphics{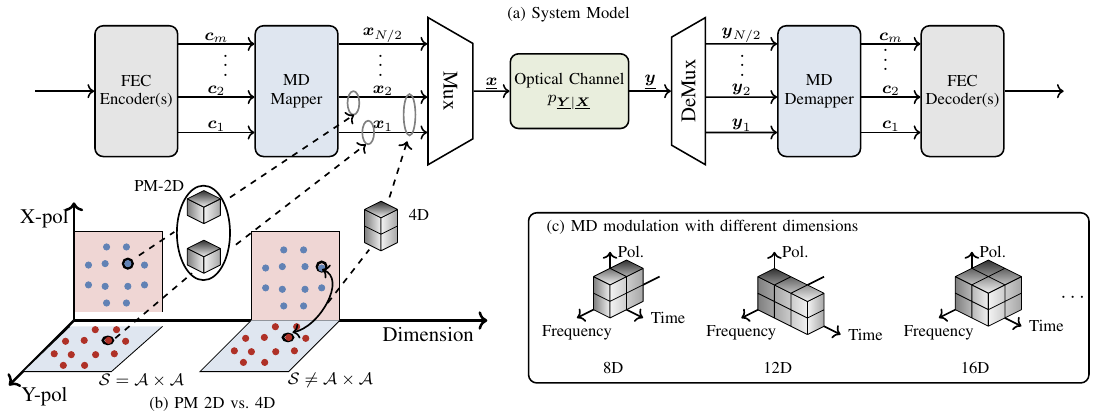}}}
\vspace{-1em}
\caption{Block diagram of CM transceiver employing MD modulation format under consideration.
The transmitter  consists of a binary FEC encoder followed by a MD mapper. The MD mapper can  either separately and  jointly generate the MD symbols $\underline{\bx}$.
Each cube denote a two-dimensional real signal space, which is  most intuitively viewed as the complex-valued symbol.
} 
\vspace{-1.3em}
\label{fig:BlockDiagram}
\end{figure*}

\section{System Model and Performance Metrics}\label{sec:model}
\subsection{System Model with Multi-Dimensional Modulation Format}
In this paper, we consider   the most popular bit-wise CM scheme for  optical communication systems, i.e., a BICM system  shown in Fig. \ref{fig:BlockDiagram}. 
The optical channel  with memory can be modeled by a conditional probability density function (PDF) $p_{\underline{\bY}|\underline{\bX}}$ with  input sequence ${\underline{\bX}}$ and output sequence  ${\underline{\bY}}$.
The transmitted symbols ${{\bX}}$ and received symbols ${{\bY}}$ in ${\underline{\bX}}$ and  ${\underline{\bY}}$ are assumed to be MD symbols with  $N/2$ complex dimensions (or equivalently, with $N$ real dimensions).
We use $k\in\{1,2,...,m\}$ to indicate  the bit position in a length-$m$ binary sequence, $j\in\{1,2,...,N/2\}$ to indicate the complex dimension and $l\in\{1,2,...,n_s\}$ to indicate the time instant.

{At the transmitter, the encoders generate coded bits
$\underline{\bc}=[\bc_1;\bc_2;...;\bc_m]$, where $\bc_k=[c_{k,1},c_{k,2},...,c_{k,n_s}]$, $k=1,2,...,m$ is the bit position and $n_s$ is the block length in symbols.
The coded bits are grouped into a length-$m$ binary sequence  and   are mapped to the  MD symbols}
\begin{align} \nonumber
\underline{\bx}=[\bx_1;\bx_2;...;\bx_{N/2}]
               =\left[
               \begin{matrix}
               x_{1,1} & x_{1,2} & \ld & x_{1,n_s}\\
               \vdots	&\vdots	& \ddots & \vdots \\
               x_{N/2,1} & x_{N/2,2} & \ld & x_{N/2,n_s}\\
               \end{matrix}
               \right],
\end{align}
where the $l$th MD symbol $[x_{1,l};x_{2,l};...;x_{N/2,l}]$ in each column is drawn from a discrete constellation    $\mathcal{S}\triangleq\{\bs_1;\bs_2;...;\bs_M\}$. 
Throughout this paper the $i$th   $N$-dimensional constellation point is denoted by $N/2$-dimensional complex symbol as  $\bs_i=[s_{i,1},s_{i,2},...,s_{i,N/2}] \in{\mathbb{C}}^{N/2}$ with $i=1,2,\ld,M$. 
{The $i$th constellation point $\bs_{i}$ is labeled by  a length-$m$ binary sequence  $\bb_i=[b_{i,1},b_{i,2},...,b_{i,m}]\in\{0,1\}^m$ from a binary labeling  set $\mathcal{B}\triangleq\{\bb_1,\bb_2,...,\bb_M\}$  using a one-to-one mapping $\bb_i\rightarrow\bs_i$.}

As shown in Fig.~\ref{fig:BlockDiagram} (a), the general structure of  MD mapper  can be used to generate arbitrary multi-dimensional  modulation formats at transmitter.  There are multiple ways to map the MD modulation to generate the optical signal. 
The two  most popular cases of 4D modulation in fiber optical communications  are shown in Fig.~\ref{fig:BlockDiagram} (b)  as  examples, 
which correspond to coherent optical communications using two polarizations of the light (each polarization includes two real dimensions, i.e., in-phase $I$ and quadrature $Q$). 
Each cube denote a two-dimensional real signal space, 
this naturally results in 4D modulation format as two cubes in Fig.~\ref{fig:BlockDiagram} (b).
In this case, the 4D modulation $\mathcal{S}$ can be  independently and separately transmitted in each complex dimension (polarization) as PM-2D, i.e., 
a single 2D modulation $\mathcal{A}$ is used to independently map information over two orthogonal polarization modes of the optical signal shown as two independent cubes.
If $\mathcal{S}\neq\mathcal{A}\times \mathcal{A}$, the  4D modulation $\mathcal{S}$  in  2$\times$2D spaces are neither identically or independently distributed, and thus, are jointly transmitted over 4D spaces  shown  as  two bonded cubes with dependency.

In  Fig.~\ref{fig:BlockDiagram} (a) and (b), we consider 4D modulation formats using $N$-dimensional space and group into  $N/4$ wavelength to form a wavelength division multiplexing (WDM) signal over the optical fiber channel as an example. Actually, the 4D  symbols (two cubes) can be also transmitted over  other $N/4$ different dimensions, e.g., time slot, spatial mode, etc.
These  degrees of freedom (wavelength, time slot and spatial mode) can  be used   to increase the number of dimensions, it can also extend 4D modulation formats to 8D, 12D, 16D and even higher dimensional modulation formats, which are shown as  multiple cubes in Fig.~\ref{fig:BlockDiagram} (c).

\vspace{-0.3em}
\subsection{Performance Metrics for BICM System}
Due to its simplicity and flexibility, BICM with SD-FEC is usually considered as an attractive option for optical fiber communication systems \cite{SmithJLT2012}, and hence, the use of the  information-theoretical performance metric GMI is preferred for coded modulation design \cite{AlvaradoJLT2018}.
Since the actual channel statistics are unknown, 
 the actual conditional distribution $p_{\underline{\bY}|\underline{\bX}}$ must be approximated  by a suboptimal  mismatched channel law $q_{\bY|\bX}$ under the assumption of memoryless and also using bit-metric decoder \cite{AlvaradoJLT2018}.
The GMI is the most popular AIR for a BICM system and  can be expressed as
\begin{align}
\label{gmi.def0}
G(\mathcal{S},\mathcal{B},q_{\bY|\bX})= \sum_{k=1}^{m} I(B_k;\bY),
\end{align}
where $\bB=[B_1,B_2,...,B_m]$ is a random vector representing the transmitted bits $\bc_l=[c_{1,l},c_{2,l},...,c_{m,l}]$ at time instant $l$, which
are mapped to the corresponding symbol $\bX_l$. 
In Eq.~\eqref{gmi.def0}, $I(B_k;\bY)$ is the mutual information (MI) between the bits and the symbols, and the notation $G(\mathcal{S},\mathcal{B},q_{\bY|\bX})$ emphasizes the dependency of the GMI on the constellation, binary labeling, and mismatched channel law. Furthermore, for any $N$-dimensional channel law, Eq. \eqref{gmi.def0} can be expressed as \cite[Eqs.~(15)--(18)]{AlvaradoJLT2018}
\begin{align}
G(\mathcal{S},\mathcal{B},q_{\bY|\bX})=&\sum_{k=1}^m \mathbb{E}_{B_k,\bY}\left[\log_2\frac{q_{\bY|B_k}(\bY|B_k)}{q_{\bY}(\bY)}\right]\\
=&m+\frac{1}{M}\sum_{k=1}^{m}\sum_{b\in\set{0,1}}\sum_{i\in\mcIkb}  \nonumber \\
\label{gmi.def}
&\hspace{-10ex}\int_{\mathbb{C}^{N/2}}q_{\bY|\bX}(\by|\bx_{i}) \log_{2}\frac{\sum_{p\in\mcIkb}q_{\bY|\bX}(\by|\bx_{p})}{\sum_{p'=1}^{M}q_{\bY|\bX}(\by|\bx_{p'})} \, \tnr{d}\by,
\end{align}
where $\mcIkb\subset\set{1,2,\ld,M}$ with $|\mcIkb|=M/2$ is the set of  indices of constellation points whose binary label is $b\in\{0,1\}$ at bit position $k$. 

The GMI  can be approximated using Monte-Carlo integration but also via Gauss-Hermite quadrature \cite{AlvaradoJLT2018}. The nonlinear optical channel is a channel with memory, 
 and thus,   the GMI Eq.~\eqref{gmi.def0} can be considered   a lower bound on the capacity of the channel \cite[Sec. V]{AlexTIT2018}. 
Monte-Carlo integration is better suited for constellations with more dimensions or when the channel law is unknown, while 
{Gauss-Hermite quadrature estimation is better-suited for numerical optimization when the channel is known analytically and it matches a quadrature.
}

For a  discrete uniformly-distributed $N$-dimensional modulation, the GMI under Gaussian noise assumption ($q_{\bY|\bX}$ is a circularly symmetric Gaussian PDF with variance $\sigma_{z}^{2}$) can be estimated via Gauss-Hermite quadrature  estimation as \cite [Eq.~(45)]{AlvaradoJLT2018}
\begin{align}\label{eq:GMI_GH}\nonumber
\text{GMI} \approx  m-  &\frac{1}{M\pi^{N/2}}\sum_{k=1}^{m}\sum_{b\in\set{0,1}}\sum_{i\in\mcIkb}\\&\sum_{l_1=1}^{\GHs}\alpha_{l_1}\sum_{l_2=1}^{\GHs}\alpha_{l_2}\cdots\sum_{l_N=1}^{\GHs}\alpha_{l_N} \cd g_i^G(\boldsymbol{\xi}),
\end{align}
with
\begin{align}\nonumber
g_i^G(\boldsymbol{\xi})=\log_{2}\frac{\sum_{p=1}^{M}\exp\left(-\frac{||\bdip||^{2}+2\sigma_{z}\Re\set{(\xi_{l_1}+\jmath\xi_{l_2})\bdip}}{\sigma_{z}^{2}}\right)}{\sum_{j\in\mcIkb}\exp\left(-\frac{||\bdij||^{2}+2\sigma_{z}\Re\set{(\xi_{l_1}+\jmath\xi_{l_2})\bdij}}{\sigma_{z}^{2}}\right)},
\end{align}
where the quadrature nodes $\xi_l$   and the weights $\alpha_l$ can be easily found (numerically) for different values of $\GHs$. 
 $\bdij \triangleq\bs_i-\bs_j$ denotes the difference between two MD symbols%,  
 and $\sigma_{z}^{2}$ is the noise variance per complex dimension.

In this paper, we use Gauss-Hermite quadrature estimation for modulation optimization with the  quadrature nodes and weights for $J=10$ in  \cite [Table III]{AlvaradoJLT2018}. 
For transmission performance evaluation, the Monte-Carlo approximation in \cite[Eq.~(31)]{AlvaradoJLT2018} is used to estimate GMI for the considered modulation formats.

\section{General Aspects of  MD Geometric Shaping} \label{sec:optimization}
\subsection{GMI-Based GS Modulation Optimization}
As shown in Eqs. \eqref{gmi.def0} and \eqref{eq:GMI_GH}, the computation of the GMI requires a joint consideration of the modulation's coordinates and its binary labeling. A GMI-based optimization under a constraint of the transmitted power $\sigma^2_x$ can be defined as 
\begin{align}\label{eq:OP_GMI}
\{\mathcal{S}^*,\mathcal{B}^*\}  = \argmax_{\mathcal{S},\mathcal{B}: E[\|\bX\|^2]\leq \sigma_x^2} G(\mathcal{S},\mathcal{B},q_{\bY|\bX}), 
\end{align}
where $\mathcal{S}^*$ and $\mathcal{B}^*$ represent the  optimal constellation and labeling solution for a given channel conditional PDF $q_{\bY|\bX}$.

Different optimization methods have been used to solve Eq.~\eqref{eq:OP_GMI}. Traditional  optimizers include for example genetic algorithm \cite{MatteoTBC2016}, pairwise optimization algorithm \cite{ZhangECOC2017} and particle swarm optimization \cite{AmirhoseinJLT2022}. 
In recent years, considering  the entire communication system design as an end-to-end reconstruction task  have received a lot of attention and the transmitter can be implemented as a  trainable GS mapper  via a  neural network  \cite{ShenLIECOC2018,RasmusECOC2018,Rasmus2019endtoend,Kadir2019endtoend,SongJSQE2022}.   
A key advantage of the neural network method is that it can be applied to arbitrary channel also  considering transceiver hardware impairments, including hardware imperfections and  nonlinear optical ones.
For more details on  end-to-end learning for optical communication systems, we refer the reader to \cite[Table I]{SongJSQE2022} and references therein. 

For  MD modulation formats,  four main factors have been identified as key factors for improving performance and complexity in the optimization:  
(\romannumeral1) the  SNR range where optimization is performed;
(\romannumeral2)  the initial  shape for a specific SNR region;
(\romannumeral3)  the bit labeling of the initial constellation;
(\romannumeral4) the number of degrees of freedom (DOFs) of the constellation to be designed. 
These factors are not very important for 1D or 2D modulation optimization, but will become crucial as the dimensionality and cardinality size increase.

\subsection{Limitations and Challenges for MD-GS  Implementation}\label{sec:limitations}

The main advantage of GS is that it only relies on the selection of the location of constellation points and the design of the corresponding  detector. 
However, some practical limitations or challenges in realistic CM schemes must be considered that can limit the applicability of GS  in optical communication systems. 
Some  drawbacks in GS schemes have been briefly discussed for example in \cite[Sec. I]{JunhoJLT2019}. 
In this section, we will discuss some limitations and challenges of MD-GS. The focus is  mainly on the  AWGN channel, however most of these limitations are also applicable to the optical channel.

\textit{1) Nonconvex multi-dimensional  optimization problem:} 
One problem often overlooked when designing MD modulation formats for  BICM system is the difficulty of  finding locations of the GS constellation points with their labeling for arbitrary channel conditions and arbitrary SEs. 
It has been shown that the optimization landscape for GMI-based optimization is highly nonconvex and  the initialization is an important design parameter, especially for the AWGN channel \cite{Kadir2019endtoend}.
Therefore, finding optimum modulation formats with good labeling based on brute force approaches quickly fails as the constellation size grows \cite{Alvarado2015_JLT}. 
It should be noted that none of the existing  approaches give guarantees on finding the global optimum. 
However, the  reported results of the optimized constellations  do outperform their counterparts (e.g., $M$QAM) in the considered  cases. 
In order to efficiently solve the nontrivial optimization problem, a fast and accurate GMI computation is crucial. 
To support an efficient MD-GS via solving the non-trivial optimization problem in Eq.~\eqref{eq:OP_GMI},  GMI estimation should be  accelerated and the search space $\{\mathcal{S}^*,\mathcal{B}^*\}$ for optimization needs to be reduced.

\textit{2) Hardware limitations:}  Another inherent assumption in most of existing GS works is that ideal  digital-to-analog converters (DACs)   and analog-to-digital converters (ADCs) are included at the transceiver. 
In order to generate and detect  high-order modulation formats, especially for geometrically-shaped modulation formats,  DACs and ADCs with a high number of resolution bits are required \cite{TimoJLT2009}. 
For { commercial quantizers with 8-bit physical resolution} and $\sim$5-6 effective number of bits (ENOB),  the use of shaped high-order modulation formats  may induce additional penalties due to the finite resolution of DAC and ADC.
It has been estimated that 20\% power reduction can be obtained when lowering the resolution from 8 to 4 bits \cite{SylvainJLT2020}.
However, GS in general induces higher peak of the driving signal's amplitude, which results in a larger quantization penalty with low resolution DAC.
Depending on the application, it is more meaningful to include finite resolution of DAC and ADC in the analysis and design hardware-friendly MD modulation formats.

\textit{3) Irregular Demapping:}  Due to the general infeasibility of Gray mapping for MD modulation formats after GS, the computational complexity of demapping symbols to soft-decision bit metrics, i.e., log-likelihood ratios (LLRs), will be increased.
To alleviate the computational complexity of LLR calculation, the well-known max-log approximation is often used and particularly interesting for the AWGN channel.  With this approximation, LLRs  for bit-wise decoding at any time instant can be computed as follows %\cite{Viterbi98}
\begin{align}\label{LLR.max-log}
\Lambda_{k}	& \approx \log\frac{\max_{\bx\in\mcIko}q_{\bY|\bX}(\by|\bx)}{\max_{\bx\in\mcIkz}q_{\bY|\bX}(\by|\bx)}.
\end{align} 

The max-log approximation therefore eliminates the exponential functions and reduces the number of Euclidean distance (ED) calculations  in a maximum likelihood (ML) optimum demapper.
For the MD AWGN channel, the max-log LLR values in Eq. \eqref{LLR.max-log} are  calculated as
\begin{align}
\label{LLR.max-log.AWGN}
\Lambda_{k}	& \approx  \frac{N}{\sigma_{\bz}^2} \left(\min_{\bx\in\mcIkz}\|\by-\bx\|^{2}-\min_{\bx\in\mcIko}\|\by-\bx\|^{2}\right).
\end{align}

For each MD received signal, the max-log demapper  needs  to calculate  all $M=2^m$ squared EDs in a $N$-dimensional space. In the case square QAM or PM-QAM modulation, the max-log approximation results in piece-wise linear relationships between the received symbol and the LLRs in 1D. This means the LLRs in this case can be calculated per real dimension rather than in $N$ dimensions.
This in turn greatly simplifies its implementation, which is partly why the max-log approximation is very popular in practice. 
For nonsquare QAM or non-Gray-labeled modulation formats, however, this is not the case.
For  most of MD modulation formats, demapper complexity could be  significantly increased  due to the larger number of  EDs calculation or larger memory requirements for pre-computed look-up table. 
In order to  reduce the computational complexity by avoiding all
possible EDs calculation, the MD modulation formats with a regular structure are preferred, e.g., lattice structures, symmetric structures and shell structures.

\section{Methods for Enabling Efficient MD Geometric Shaping and Its Implementation}\label{sec:solutions}
In this section, 
four potential methods of overcoming the limitations discussed in Sec.~\ref{sec:limitations}  are shown. 
The first two methods aim to accelerate the GMI approximation while keeping high accuracy (\textit{Example~1} and \textit{2}). 
Next, symmetric constraints are added to simplify the optimization and to reduce the hardware penalty (\textit{Example~3}). 
Finally,  performance metrics for assessing modulation-dependent NLI are discussed (\textit{Example~4}).

\begin{figure}[!tb]
    \centering
 {\includegraphics{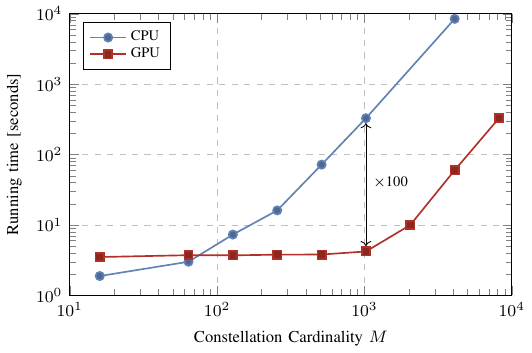}}
   \vspace{-1.9em}
   \caption{Comparison of GMI estimation running time  between  CPU and  GPU  for 4D modulation formats via Gauss-Hermite quadrature (Eq. \eqref{eq:GMI_GH}). Results were obtained on a workstation with an Intel Core i7-11700 CPU and a NIVIDA GeForce RTX 3090 GPU, respectively.}
    \label{fig:cpugpu}
       \vspace{-1em}
\end{figure}

\begin{example}[GPU-acceleration for  GMI estimation] 
It is known that GMI-based optimization of large constellations  with high dimensionality (i.e., the top right corner of Fig.~\ref{fig:MD_gains}) is computationally demanding. 
{This has been considered as a challenging issue for such large constellations since the exact MI and GMI estimation requires enumerating all constellation points (\textit{Example 2} will discuss a  weighted sampling method to avoid enumerating all the points.)}.
{To solve this computation issue, parallel GPU computing can be used to accelerate the computational speed of GMI by parallelizing the quadrature approximation in Eq.~\eqref{eq:GMI_GH}. 
Fig.~\ref{fig:cpugpu} shows a comparison of  GMI computation running time between CPU and GPU for 4D modulation formats with different cardinatity size $M =\{ 16,64,128,256,512,1024,2048,4096,8192\}$.
The simulation was obtained with using an  Intel Core i7-11700 CPU and an NIVIDA GeForce RTX 3090  GPU via Python. 
The results shown in Fig.~\ref{fig:cpugpu} indicate that the GPU-based GMI estimation is at least 100 times faster than the CPU-based when the constellation cardinality size larger than $10^3$.
 {Note that the running time of GPU-based GMI estimation also increases due to the parallelization degree limitation of the used GPU when the  constellation cardinality is very large. Thus, larger speed-ups are expected when using more powerful GPUs or multi-GPUs architecture.}
Overall, this acceleration can greatly simplify to make the MD constellation optimization to be possible.
} 
\end{example}

In the following example, the GMI estimation for very large constellations  is particularized to the relevant case of Monte-Carlo integration.

\begin{example}[Low complexity GMI estimation for very large constellations]%
To numerically estimate Eq. \eqref{gmi.def}, the key point is to calculate $q_{\bY}(\by)=\frac{1}{M}\sum_{\bx\in\mathcal{S}}q_{\bY|\bX}(\by|\bx)$, which is infeasible when $M$ is very large.
Even though Monte-Carlo techniques can somewhat   estimate the GMI faster than the Gauss-Hermite method,  they can be very inaccurate when the total number samples are limited, as shown in \cite[{Example}~4.26]{Szczecinski2015BICM}.
One way is to use  a weighted sampling method, called  importance sampling, which oversamples from the important region, thus making Monte-Carlo accurate \cite{ShenLiTCOM2021}. 
The intuition behind  importance sampling is that only a fraction of all constellation points  contribute significantly to the sum in Eq. \eqref{gmi.def}.
By following the method in \cite[Eqs.~(22)--(23)]{ShenLiTCOM2021}, GMI estimation base on Monte-Carlo method can be simplified but also accurate by choosing a proper  important constellation set.
{For constellation with good structure, e.g., Voronoi constellation with around  $10^{46}$ 32-dimensional lattice  points, only $10^5$ random samples for each important subset are needed for accurate GMI estimation.
} 
\end{example}

The  discussed  methods in \textit{Example 1} and \textit{Example 2} show that the  optimization  in Eq.~\eqref{eq:OP_GMI} for  high-dimensional  modulation formats can be  made significantly more efficient by accelerating the GMI estimation.
However, the search space for large constellations and/or for constellations with high dimensionality is still  computationally very demanding due to the large continuous search space. 
In addition, potentially irregular formats obtained after optimization also impose strict requirements on the generation and detection of the signals, due to the need of high-resolution DACs and ADCs.
To solve the multi-parameter optimization challenges of MD-GS and also to  achieve a good performance-complexity trade-off, constraints such as constant modulus \cite{Kojima2017JLT,BinChenJLT2019}, orthant-symmetry (OS) \cite{BinChenJLT2021}, X-Y symmetry and 4D shells \cite{Sebastiaan2022arxiv,Sebastiaan2022arxiv_JLT} have been proposed to design MD formats. 
These solutions have shown a small performance loss with respect to the unconstrained optimizations in AWGN channel (or linear region of optical channel), but can achieve even better performance in the nonlinear optical fiber channel.

\begin{example}[Orthant-symmetry constraint]
For an $N$-dimensional constellation $\mathcal{S}$ with $M=2^m=|\mathcal{S}|$ points, each point in $\mathcal{S}$ has four DOFs. Thus, geometrically optimizing a 4D constellation in an unconstrained way results in $N\cdot  2^m$ DOFs, which quickly becomes challenging as $N$ and $m$ increase.
To  reduce the  number of DOF for MD geometric shaping and also to reduce
the transceiver requirements, OS constraint has been proposed   in \cite{BinChenJLT2021} to  reduce the dimensionality of searching space within the first orthant.
Orthant-symmetric labeled constellations
can be generated from any first-orthant constellation,
where the constellation points are obtained
by folding the first-orthant points to the remaining
orthants. 
For a more detailed description of these concepts, we refer the reader to \cite{BinChenJLT2021}.
In addition, a recent work combining  OS  with X-Y symmetry and 4D shells  constraints has been investigated  in the context of 400ZR standard \cite{Sebastiaan2022arxiv,Sebastiaan2022arxiv_JLT}.

By applying the OS constraint, the optimization problem in Eq. \eqref{eq:OP_GMI}   can be simplified  as 
\begin{align}\label{GMI-optimal-OS}
\{\mathcal{S}^*_1,\mathcal{B}^* _{1}\}=&\argmax_{\mathcal{S}_1,\mathcal{B}_1:E[||X||^2]\leq \sigma^2_x} \{G(\mathcal{S},\mathcal{B},q_{\bY|\bX})\}
\end{align}
where  $\mathcal{S}_1  \subset \mathcal{R}^{+}$ and $\mathcal{B}_1$ denote the first-orthant constellation points and their corresponding labelings. The obtained solution  $\{\mathcal{S}^*_1,\mathcal{B}^*_1\}$ can  be used to obtain the complete labeled constellation $\{\mathcal{S},\mathcal{B}\}$.
Therefore, the OS constraint can reduce the amount of DOFs by a factor of $2^N$.

{The insets of Fig. \ref{fig:OS_constellations} show 2D projections of two AWGN-optimized constellations with $N=4$ and $M=128$  }, where inset (a) is  obtained via Eq. \eqref{eq:OP_GMI} (called 4D-GS128), and inset (b) is optimized with an OS constraint by solving Eq. \eqref{GMI-optimal-OS} (called 4D-OS128).
The amount of DOFs is reduced from $4\cdot 2^7=2^9$ to $4\cdot 2^7/2^4=32$.
As shown in \cite{WeiICOCN2021}, the optimization with OS constraint is much faster to be converged and potentially avoid a suboptimal result with respect to the optimization without OS constraint.
In addition, the  constellation in Fig.~\ref{fig:OS_constellations} (a) is not well structured, which could lead to a high penalty when using  high-speed DAC with limited ENOB. 
Therefore, adding a proper constraint, e.g. OS,  does not only increase  optimization speed,  but also reduces the DAC/ADC resolution requirements.

\begin{figure}[!tb]
  \centering
%  {\input{./tikz/Performance_DAC.tikz}}
 {\includegraphics{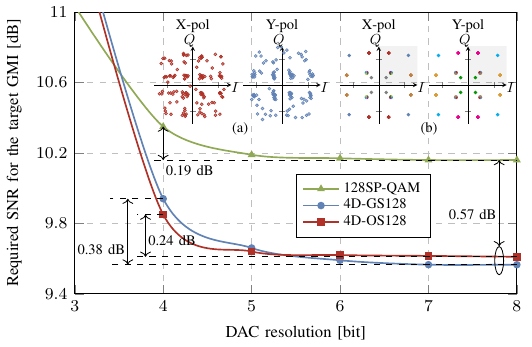}}
\vspace{-2em}
\caption{Required SNR vs.  DAC resolutions for  three 4D 128-ary modulation formats at GMI of 5.95~bit/4D-sym. Insets: (a) optimized via Eq. \eqref{eq:OP_GMI} as 4D-GS128, (b) optimized via Eq. \eqref{GMI-optimal-OS} as 4D-OS128. 
} 
\vspace{-1em}
\label{fig:OS_constellations}
\end{figure}

In order to compare the impact of finite-resolution DACs, three modulation formats (128SP-QAM, 4D-GS128 and 4D-OS128) are evaluated. At the transmitter, the symbol sequence at 2 samples per symbol was filtered with 
pulse shaping filter, and then  the driving signals are linearly quantized with a variable number of quantization levels for DAC. {The effect of the ADCs are not considered in this example.}
To measure the impairments of limited DAC resolution, we measured the minimum SNR required to achieve the target GMI.   
The results of the numerical simulations  at a symbol rate of 45~GBd over AWGN channel are shown in Fig.~\ref{fig:OS_constellations}. 
It shows the required SNR at GMI  of 5.95~bit/4D-sym as function of the DAC resolution for three modulation formats.
When comparing the performance at high DAC resolutions, the two 4D shaped modulation formats can provide around a gain of 0.57~dB.
{We can also observe  that decreasing the DAC resolution below four bits results in significantly penalties for all modulation formats.
Despite the small gains provided by 4D-GS128  at high DAC resolutions, when the DAC resolution is reduced to 4, the performance of 4D-GS128 (0.38~dB penalty) degrades more rapidly than 4D-OS128 (0.24~dB penalty)  and 128SP-QAM (0.19~dB penalty) due to the larger quantization distortion for the neighboring points.} 
\end{example}

\begin{example}[{Performance metrics for nonlinearity-tolerant 4D modulation}] 
Fig.~\ref{fig:NLI_KPI} shows NLI-related performance metrics versus iteration number in the optimization process for  two sets of 4D modulation formats: AWGN-optimized (blue lines) and NLI-optimized (red lines). AWGN-optimized modulation formats are obtained by maximizing GMI for a fixed SNR, while NLI-optimized formats are obtained by maximizing GMI with considering the NLI power to trade-off between shaping and nonlinearity in nonlinear channel. 
To evaluate the NLI effect of the optimized modulation formats in the optical channel, we consider a dual-polarized, single channel waveform  over a 234~km single span  SSMF. 
{More details about the  nonlinearity-tolerant 4D modulation optimization are given in Sec.~\ref{sec:single_span}.}
By  treating the modulation-dependent  NLI as Gaussian noise,  the change in modulation by moving the 4D symbols during the optimization can be reflected in effective SNR as shown in Fig.~\ref{fig:NLI_KPI}.

\begin{figure}[!tb]
\centering
\input{./pdf/NLIvsIter.tex}
\vspace{-1.5em}
\caption{NLI-related performance metrics vs. the 4D modulation optimization process  for optical fiber simulations of single channel with  symbol rate of 45~GBaud over single span of 234~km single-mode fiber (see Sec.~\ref{sec:single_span} for details of optimization). The 4D symbol sequences are  normalized to $E[||\bX||^2]=2.$ to compute the performance metrics.} 
\label{fig:NLI_KPI}
\vspace{-1em}
\end{figure}
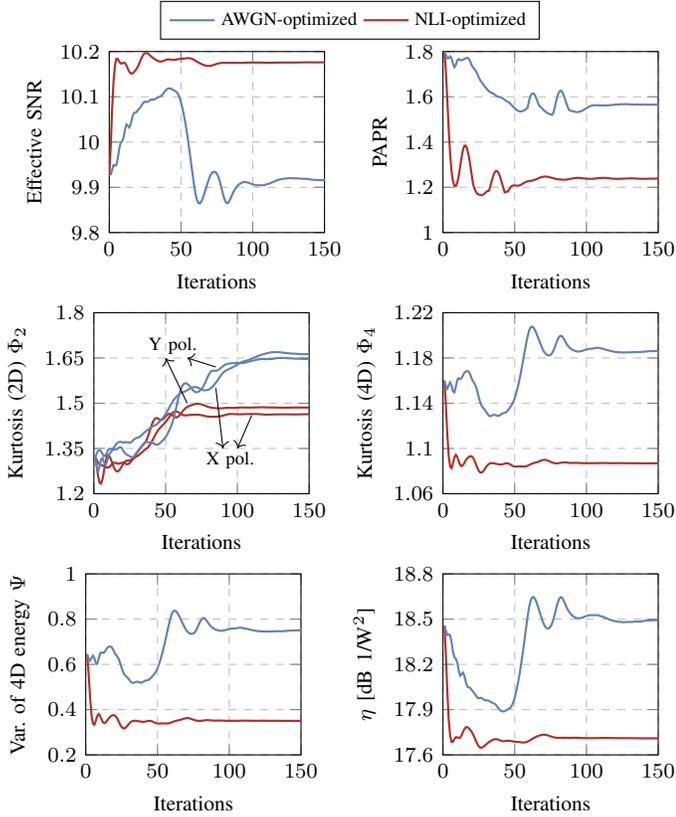

In Fig.~\ref{fig:NLI_KPI}, the nonlinear tolerance related performance metrics for showing variations of symbol
energies, i.e., peak-to-average power ratio (PAPR), kurtosis (here, we use  $\Phi_2$ and $\Phi_4$ to denote the kurtosis of 2D symbols and 4D symbols, respectively),  the variance of  the transmitted 4D symbols’ energy $\Psi$  and  the NLI power coefficient $\eta$ (see \cite[Eq. (1)]{GabrieleOFC2021}) are also shown for each optimized 4D formats during the iterative optimization.\footnote{Note that the variance of  the transmitted 4D symbols’ energy is the deviation of the signal energy from the mean value and is also determined by the kurtosis of the 4D transmitted symbol as  $\Psi=\mathbb{E}[||\bX||^2](\Phi_4-1)$ \cite[Eq. (48)]{KaiquanJLT2021}.  {Similar metrics with considering sliding window were defined in \cite{KaiquanJLT2021,JunhoJLT2022} to evaluate probabilistic shaping with  time-varying correlations.}}

By comparing the optimization traces between the AWGN-optimized (blue lines) and NLI-optimized formats (red lines), we can observe that an NLI-tolerant 4D modulation format (higher effective SNR and smaller $\eta$)  has  in general  a smaller PAPR, smaller 4D kurtosis and smaller variance of  the transmitted 4D symbols’ energy, which highlights  the fact that all these performance metrics are inherently related to  NLI. 
{However, PAPR only partially reflects the energy variation of symbols  (depends on the few constellation points) and 2D kurtosis $\Phi_2$   neglects the 4D geometry (only  takes  the 2D symbol energy into account).
From the  optimization process shown in Fig.~\ref{fig:NLI_KPI}, we can see that PAPR and $\Phi_2$ can not  reflect well the trend of modulation-dependent nonlinear tolerance (effective SNR and $\eta$) for dual-polarization transmission systems, and thus cannot reflect the complete nonlinear performance.
These two performance metrics  can  only be regarded as a rough indication of nonlinearity tolerance of the  4D  formats.} 
In contrast, Kurtosis of 4D symbols  $\Phi_4$ and the variance of 4D symbol energy  $\Psi$ are able to capture the partial  effect of dual-polarizations symbol statistics on the NLI, and thus, they are  in general consistent with the trend of effective SNR and NLI power coefficient $\eta$, but they are not exactly matching. 
Therefore,  we  conjecture that  $\Phi_4$ or $\Psi$  can be  also considered as efficient performance metrics for designing NLI-tolerant 4D modulation formats by avoiding the complex perturbation approach for calculating the average NLI power coefficient $\eta$. 
The coefficient $\eta$ for DP-4D formats has been derived in  \cite{GabrieleEntropy2020} and demonstrated to predict the NLI in optical systems in \cite{GabrieleOFC2021}. 
\end{example}

{We conclude this section by emphasizing that the four examples above showed that adding proper constraints in the optimization can reduce the  transceivers complexity and penalty for implementation MD-GS modulation formats.
Such constraints could allow us to  design higher SE modulation formats and introduce specific correlations between different dimensions  as the number of dimension increases.}
{As we will see in the next two sections, optimized 4D modulation formats and its optimized labeling provide an excellent  transmission performance with respect to conventional QAM and SP-QAM modulation formats in optical transmission system.}

\remark{All of these existing optimized modulation formats obtained with  constraints have not been proven to be the best for either the  AWGN channel or  the  optical fiber channel for the BICM paradigm.
Finding the optimum modulation formats remains an open research problem. 
Despite this cautionary statement, the methods discussed above are known to perform well  based on bit-wise decoders in terms of GMI, as shown in Sec.~\ref{sec:multi_span} and Sec.~\ref{sec:single_span}.
}

% two sections of results

\input{Section_MultiSpan}
\input{Section_SingleSpan}

\section{Conclusions}\label{sec:con}
In this paper, we  reviewed the existing multi-dimensional geometrically-shaped modulation formats for fiber optical communication systems and showed that they could provide potentially larger gains.
The key challenges and different methods for overcoming the  limitations  associated with the develop of  MD-GS formats were presented
and discussed.
The main focus of this paper was on the 4D modulation formats for
dual-polarization optical fiber channel.
With the aid of adding efficient  constraints for optimizing modulation formats, we numerically assessed a series of 4D modulation formats via numerical simulation. 
We showed that the 4D-optimized modulation formats can be a solution for multi-rate applications between 5 and 10~bit/dual-pol with {a transmission reach extension of up to 25\% in multi-span systems.}
In addition, up to 0.25~dB NLI gains in terms of effective SNR are demonstrated for NLI model-optimized modulation  over  regular 4D format and Gaussian channel-optimal 4D format in a single-span transmission system. 
 The results in this work confirm that the multi-dimensional modulations could be a good alternative for high capacity  transmission systems and offer  substantial potential gains in the nonlinear optical fiber channel.

Note that the performance of MD modulation also depends on which dimensions are selected. 
The methods described in this paper can be extended to nonlinear optical channel by jointly combing  other dimensions with different correlated property and also to  channels with memory. {This is left for further investigation.}

All the analysis presented in this paper was only  considered bitwise decoder (i.e., BICM) for  multi-dimensional geometric shaping. 
Better performance are expected if a hybrid CM with combing of BICM and multi-stage decoding are used. 
In these cases, we conjecture the  a performance metric of combing MI and GMI to be the correct metric to design MD modulation formats.  
This investigation and comparison are interesting future research avenues.

\balance
%%%%%%%%%%%%%%%%%%%%%%%%%%%%%%%%%%%%%%%%%%%
%%              References               %%
%%%%%%%%%%%%%%%%%%%%%%%%%%%%%%%%%%%%%%%%%%%
\bibliographystyle{IEEEtran}
\bibliography{references_4D64PRS,references}
\end{document}

%% file: myFigures.tex
\usetikzlibrary{plotmarks,matrix,chains,scopes,fit,calc,shapes,positioning,decorations,intersections,arrows,backgrounds,shadows}
\usetikzlibrary{fit}
\usetikzlibrary{shapes.multipart}
\usetikzlibrary{positioning}
\usetikzlibrary{shapes}
\usetikzlibrary{arrows.meta}
\newlength\FigureHeight
\newlength\FigureWidth

%% file: myStyles.tex
\definecolor{myDarkGreen}{rgb}{0.00000,0.58824,0.00000}%
\definecolor{uniform}{rgb}{0.00000,0.58824,0.00000}%
\definecolor{AWGNreference}{rgb}{0.00000,0.58824,0.00000}%

%% file: myMacros.tex
\DeclareMathOperator*{\argmax}{argmax}

\newcounter{lemma}

\newtheorem{exampleplain}{Example}
\newtheorem{remark}{Remark}
\newenvironment{example}{\begin{exampleplain}}{~\hfill$\vartriangle$\end{exampleplain}}

\newcommand{\cd}{\cdot}
\newcommand{\mcIkb}{\mathcal{I}_{k}^{b}}
\newcommand{\mcIkz}{\mathcal{I}_{k}^{0}}
\newcommand{\mcIko}{\mathcal{I}_{k}^{1}}
\newcommand{\bdij}[0]{\boldsymbol{d}_{ij}}

\newcommand{\bdip}[0]{\boldsymbol{d}_{ip}}

\newcommand{\set}[1]{\{#1\}}

\newcommand{\ld}{\ldots}

\newcommand{\bc}{\boldsymbol{c}}
\newcommand{\bB}{\boldsymbol{B}}\newcommand{\bb}{\boldsymbol{b}}
\newcommand{\bX}{\boldsymbol{X}}\newcommand{\bx}{\boldsymbol{x}}

\newcommand{\bY}{\boldsymbol{Y}}\newcommand{\by}{\boldsymbol{y}}
\newcommand{\bz}{\boldsymbol{z}}

\newcommand{\bs}{\boldsymbol{s}}

\newcommand{\GHs}{J}

\newcommand{\tnr}[1]{{\textnormal{#1}}}

\definecolor{RED}{rgb}{1,0,0}

%% file: myPackages.tex
\usepackage[utf8]{inputenc}

\usepackage{amsmath,amssymb}

\usepackage{xcolor}
\usepackage{graphicx}
\usepackage{tikz}
\usepackage{pgfplots}
\usepackage{float}
\usepackage{balance}
\usepackage{pgfplotstable}
\usepackage{xspace}
\usepackage[normalem]{ulem}
\usepackage[pdftex]{hyperref} %pdflatex
\hypersetup{colorlinks,linkcolor=red,citecolor=blue} % hyperref colors

\usepackage{dsfont,empheq}
\usepackage{multicol,multirow}

\usepackage{cite}

%% file: acronyms.tex
\newacronym{KK}{KK}{Kramers-Kronig}
\newacronym{CSPR}{CSPR}{carrier-to-signal power ratio}
\newacronym{KKRX}{KKRx}{Kramers-Kronig Receiver}
\newacronym{SSBI}{SSBI}{signal-signal beat interference}

\newacronym{DSP}{DSP}{digital signal processing}
\newacronym{MIMO}{MIMO}{multiple-input multiple-output}
\newacronym{TDE}{TDE}{time domain equalizer}
\newacronym{FDE}{FDE}{frequency domain equalizer}
\newacronym{LMS}{LMS}{least means square}
\newacronym{DDLMS}{DD-LMS}{decision directed least means square}
\newacronym{FFE}{FFE}{feed-forward equalizer}
\newacronym{FBE}{FBE}{feedback equalizer}

\newacronym{SMF}{SMF}{single-mode fiber}
\newacronym[plural=SSMFs]{SSMF}{SSMF}{standard single-mode fiber}
\newacronym[plural=FMFs]{FMF}{FMF}{few-mode fiber}
\newacronym{FMF12}{FMF12}{12 mode FMF}
\newacronym{MMF}{MMF}{multi-mode fiber}
\newacronym{SI}{SI}{step index}
\newacronym{GI}{GI}{graded index}
\newacronym{DCF}{DCF}{dispersion compensated fiber}

\newacronym{SDM}{SDM}{space division multiplexing}
\newacronym{MDM}{MDM}{mode division multiplexed}
\newacronym{WDM}{WDM}{wavelength division multiplexing}

\newacronym{LP}{LP}{linear polarized}
\newacronym[plural=MMUXs,firstplural=mode multiplexers]{MMUX}{MMUX}{mode multiplexer}
\newacronym{PL}{PL}{photonic lantern}
\newacronym{3DWG}{3DWG}{3D-waveguide}
\newacronym{MDL}{MDL}{mode dependent loss}
\newacronym{DGD}{DGD}{differential group delay}
\newacronym{DMGD}{DMGD}{differential mode group delay}
\newacronym{QSM}{QSM}{quasi-single-mode}
\newacronym{GIMMF}{GI-MMF}{graded-index multi-mode fiber}

\newacronym{SSB}{SSB}{single side band}
\newacronym{QPSK}{QPSK}{quadrature phase shift keying}
\newacronym{QAM}{QAM}{quadrature amplitude modulation}
\newacronym{RRC}{RRC}{root-raised-cosine}

\newacronym{ECL}{ECL}{external cavity laser}
\newacronym{CW}{CW}{continuous wave}
\newacronym[plural=DFBs]{DFB}{DFB}{distributed feedback laser}

\newacronym[plural=DACs]{DAC}{DAC}{digital to analog converter}
\newacronym{ADC}{ADC}{analog to digital converter}
\newacronym{PRBS}{PRBS}{pseudo-random bit sequence}
\newacronym{LO}{LO}{local oscillator}
\newacronym{EDFA}{EDFA}{erbium doped fiber amplifier}
\newacronym{MZM}{MZM}{Mach-Zehnder modulator}
\newacronym{DP-MZM}{DP-MZM}{dual-polarization Mach-Zehnder modulator}
\newacronym{ChUT}{ChUT}{channel under test}
\newacronym{WSS}{WSS}{wavelength selective switch}
\newacronym[plural=VOAs]{VOA}{VOA}{variable optical attenuator}
\newacronym[plural=PDCRXs]{PDCRX}{PDCRX}{polarization diverse coherent receiver}
\newacronym{DSO}{DSO}{digital storage oscilloscope}
\newacronym{ASE}{ASE}{amplified spontaneous emission}
\newacronym{PBS}{PBS}{polarization beam splitter}
\newacronym{PD}{PD}{photodiode}

\newacronym{OSNR}{OSNR}{optical signal to noise ratio}
\newacronym{BER}{BER}{bit error rate}
\newacronym{IL}{IL}{insertion loss}

\newacronym{SDFEC}{SD-FEC}{soft decision forward error correction}
\newacronym{HDFEC}{HD-FEC}{soft decision forward error correction}
\newacronym{FEC}{FEC}{forward error correction}

\newacronym{AIR}{AIR}{achievable information rate}
\newacronym{AR}{AR}{achievable rates}
\newacronym{MI}{MI}{mutual information}

\newacronym{OVNA}{OVNA}{optical vector network analyzer}
\newacronym{NIR}{NIR}{near infrared}
\newacronym{CD}{CD}{chromatic dispersion}
\newacronym{OTDR}{OTDR}{optical time domain reflectometry}
\newacronym{OFDR}{OFDR}{optical frequency domain reflectometry}
\newacronym{GPU}{GPU}{graphics processing unit}
\newacronym{SVD}{SVD}{singular value decomposition}
\newacronym{WGN}{WGN}{white Gaussian noise}
\newacronym{AWGN}{AWGN}{additive white Gaussian noise}
\newacronym{PDL}{PDL}{polarization dependent loss}
\newacronym{SPS}{sps}{samples-per-symbol}

%% file: colors.tex
\definecolor{blue}{rgb}{0.38, 0.51, 0.71} %glaucous, 97,130,181, #6182B5
\definecolor{darkblue}{RGB}{17, 42, 60} % 112A3C
\definecolor{red}{RGB}{175, 49, 39} % AF3127

\definecolor{orange}{RGB}{217, 156, 55} % D99C37
\definecolor{green}{RGB}{144, 169, 84} % 90A954
\definecolor{palegreen}{RGB}{197, 184, 104} % C5B868

\definecolor{yellow}{RGB}{250, 199, 100} % FAC764
\definecolor{brokenwhite}{RGB}{218, 192, 166} % DAC0A6
\definecolor{brokengrey}{rgb}{0.77, 0.76, 0.82} % {196,194,209}, C4C2D1

%% file: table/MD_GS_review.tex
\scriptsize

\resizebox{\textwidth}{!}{

\begin{tabular}{@{}@{\hskip 0.2ex}c@{\hskip 0.3ex}@{\hskip 0.3ex}c@{\hskip 0.3ex}@{\hskip 0.3ex}c@{\hskip 0.2ex}@{\hskip 0.2ex}c@{\hskip 0.2ex}@{\hskip 0.2ex}c@{\hskip 0.1ex}@{\hskip 0.1ex}c@{\hskip 0.2ex}@{\hskip 0.2ex}c@{\hskip 0.5ex}@{\hskip 0.5ex}c@{\hskip 0ex}@{}}
			%{ccccccc}
			\hline	
			
			\hline 
			%Ref. 	
				&Modulation	& SE &	Channel		& Baseline &Performance metric& Gain 		&  Description\\
			\hline 
			
			\hline		
			\multicolumn{8}{c}{\textbf{2D-GS}}\\
			\hline
		%	\cite{BinECOC2018}
			&GS-16 \cite{BinECOC2018}	&4~bit/2D-sym	& AWGN	         	& 16QAM 	&MI/GMI	&  SNR: 0.1--0.2~dB &    Optimize for multiple SNRs   \\
		%	\cite{ZhangECOC2017}
			&GS-32 \cite{ZhangECOC2017}	&5~bit/2D-sym	& AWGN/Experiments	& 32QAM 	&GMI	&SNR: {$0.5$~dB} &  SNR=11~dB\\
		%	\cite{BinICTON2018}
			&GS-64 \cite{BinICTON2018}	&6~bit/2D-sym	&AWGN/SSFM	        & 64QAM 	&MI/GMI	& SNR: 0.5--0.6~dB 	&      250~km reach $\uparrow$ (+12\%) \\
			&GS-64 \cite{MariaJLT2020}	& 6~bit/2D-sym	&  Experiments		& 64QAM 	&GMI	&SNR: {$0.5$~dB} 	 	& 74.38~Tbit/s in  C+L band\\
			&GS-128  \cite{SchaedlerOFC020}	&7~bit/2D-sym	& Experiments	& 128QAM 	&BER	& SNR: 1--1.2~dB 	 	&  800~Gb/s, 1~Tb/s\\
		%	\cite{BinECOC2018,RasmusECOC2018}
			&GS-256 \cite{BinECOC2018,Rasmus2019endtoend,Kadir2019endtoend} &	8~bit/2D-sym	& AWGN/SSFM		& 256QAM 	&MI/GMI		&SNR: $0.8$~dB 	 	& 10\%-17\% reach $\uparrow$ \\
			&IPM-256 \cite{LotzJLT2013} &	8~bit/2D-sym	& Experiments	& 256QAM 	&MI	&SNR: $0.88$~dB 	 	&  BICM-ID, 800~km (single channel)\\
		%	\cite{Kadir2019endtoend}
			&GS-1024 \cite{Kadir2019endtoend}&	 10~bit/2D-sym	& AWGN		& 1024QAM 	&GMI		&SNR: 1~dB 			& 26\% reach $\uparrow$\\
			%\cite{XX}		&XX		& AWGN		&  $\infty$QAM 	&MI		& 1.53~dB 			& XX\\
			&GS-64/256/1024 \cite{GaldinoPTL2020} &	 6--10~bit/2D-sym	& Experiments		& 64/256/1024QAM 	&GMI		& SNR: 0.5--1~dB
			 & 18\% throughput $\uparrow$, 178~Tbit/s in S+C+L band\\
	    	&GS2D/GSNL2D \cite{Sillekens2022arxiv} &	 3--13~bit/2D-sym	& AWGN/SSFM		& QAM 	&GMI		& SNR: close to 1.53~dB
			 & Optimize for multiple SNR\\
			\hline
			\multicolumn{8}{c}{\textbf{4D-GS}}\\
			\hline
			&PS-QPSK \cite{Karlsson:09,AgrellJLT2009}	&3~bit/4D-sym	& AWGN		&PM-QPSK	& BER		& SNR: 0.97~dB 	&  at BER of $10^{-3}$, SE:  25\% $\downarrow$ w.r.t QPSK\\
			&8PolSK–QPSK \cite{Chagnon:13}	&5~bit/4D-sym	& Experiments		&PM-8QAM 	& BER		& SNR: 0.5~dB 			&  34\% reach $\uparrow$, 3800~km\\
			&64SP-12QAM \cite{NakamuraECOC2015}	&6~bit/4D-sym	& Experiments		&PM-8QAM 	& BER		& SNR: 0.3~dB 			&  8\% reach $\uparrow$ (3 channels)\\
		%	\cite{Kojima2017JLT}
			 &4D2A8PSK \cite{Kojima2017JLT}	& 5--7~bit/4D-sym	& SSFM		&PM-8QAM/32SP-QAM	& GMI		& SNR: 0.6--2.2~dB 			&  4D constant-modulus, DM link\\
		%	\cite{BinChenJLT2019}
			&\multirow{2}{*}{4D-64PRS \cite{BinChenJLT2019,SjoerdOECC2019}	} & \multirow{2}{*}{6~bit/4D-sym }	& AWGN  		& \multirow{2}{*}{PM-8QAM }		&\multirow{2}{*}{GMI}	& \multirow{2}{*}{SNR: 0.7--0.9~dB} 			&  4D constant-modulus\\
		%	\cite{SjoerdOECC2019}
			       &  	&	& SSFM/Experiments		& 			&  		& 	 &  %EDFA/Raman,
			       16\% reach $\uparrow$, 11700~km (11 channels)\\
			%   \cite{ErikssonOE13,KashiECOC2015} 		&SP128-QAM		&SSFM		& PM-16QAM&	BER		& xx~dB 	 			& 7~bit/4D-sym, 比较不公平\\
			&4D-OS128 \cite{BinChenJLT2021}	&7~bit/4D-sym	& SSFM/Experiments			& 128SP-QAM &	GMI 		& 
			SNR: 0.65~dB
				&  %EDFA/Raman,
				15\% reach $\uparrow$, 9000~km (11 channels)\\
    &{4D-GSS \cite{Sebastiaan2022arxiv,Sebastiaan2022arxiv_JLT}}&	{8~bit/4D-sym}	& {SSFM}				& {PM-16QAM} &	{MI/GMI} 		&{SNR: 0.23--0.27~dB} 			&   {3\% reach $\uparrow$, single span 160~km  (single channel)}\\
			&4D-512-Hurwitz \cite{FreyJLT2020}&	9~bit/4D-sym	& AWGN/Experiments				& PM-16QAM &	MI/GMI 		&SNR: 0.8~dB 			&   Two-stage demodulation, 400G over 120~km\\
			&$C^*_{64/128/256/512}$  \cite{GabrieleOFC2022} &	 6--9~bit/4D-sym	& SSFM		&PM-$M$QAM/4D-OS128 &MI		& SNR: {0.2--0.7~dB}
			 & 4\%--13\%  reach $\uparrow$ (sinlge channel)\\
		  %&4D-1024 \cite{Vinicius2021arXiv}&	10~bit/4D-sym	& SSFM			& PM-32QAM &	GMI 		& SNR: \todo{0.7~dB}			& 13.6\% reach $\uparrow$ (5 channels)\\
			&GS4D-4096 \cite{Sillekens2022arxiv}&	12~bit/4D-sym	& AWGN/SSFM & PM-64QAM &	GMI 		&SNR: 0.5--0.8~dB 			&   10\% reach $\uparrow$, 1500~km (11 channels)\\
			\hline
		\multicolumn{8}{c}{\textbf{8D-GS}}\\
			\hline
			\multirow{2}{*}{
			%	\cite{Shiner:14}
			}
			&\multirow{2}{*}{X-constellation \cite{Shiner:14}}	&  \multirow{2}{*}{4~bit/8D-sym}	& AWGN		&  \multirow{2}{*}{PM-BPSK} &	BER		& SNR: 0.53~dB 			& at  BER of $3.5\times10^{-2}$\\
			& 	&	& Experiments		&    &	Q-factor		&Q-factor: 2~dB 			& {DM link, 5000~km}\\
		%		\cite{Bendimerad:18}
			&\multirow{2}{*}{PB-5b8D, PA-7b8D \cite{Bendimerad:18}}	&5~bit/8D-sym	& SSFM		&  PM-BPSK &	Q-factor		& Q-factor: 0.1~dB 			&  SE: 25\% $\uparrow$ w.r.t BPSK \\
			& &7~bit/8D-sym	& SSFM		&  PM-QPSK &	Q-factor		& Q-factor: 0.8~dB 			&  SE:  12.5\% $\downarrow$ w.r.t QPSK \\
			&PB-PS-QPSK \cite{El-RahmanJLT2018}	&6~bit/8D-sym	& SSFM			& PS-QPSK &	MI 		&  	Rate: 1.5~Gb/s &  DM link, 10000~km, 35~GBd\\
            &8b-8D-sphere \cite{KoikeAkinoECOC2013}	&8~bit/8D-sym	& SSFM			& PM-QPSK &	BER 		&  	SNR: 0.2--1~dB &   at BER of  $10^{-2}$ and $10^{-3}$\\				
   %	\cite{BinChenPTL2019}
			&\multirow{2}{*}{8D-PRS2048 \cite{BinChenPTL2019,SjoerdECOC2019}}&	\multirow{2}{*}{11~bit/8D-sym}	& AWGN+SSFM		&  \multirow{2}{*}{PM-8QAM/4D2A8PSK} &	 \multirow{2}{*}{GMI}	&	\multirow{2}{*}{SNR: 0.25--0.7~dB} 			&5\%--15\% reach $\uparrow$ \\
		%	\cite{SjoerdECOC2019}
			&	&	& Experiments		&   &	 	&  			&EDFA (11 channels)\\
			\hline
			\multicolumn{8}{c}{\textbf{MD-GS}}\\
			\hline
			&\multirow{2}{*}{12D \cite{ReneOFC2020}}	&\multirow{2}{*}{12~bit/12D-sym}	& AWGN			& \multirow{2}{*}{PM-QPSK} &	BER 		&  	SNR: 1.2~dB		&  at BER of $10^{-4}$\\
			&	&	& Experiments			&  &	MI 		& Reach: 8\%--15\% 			&  MCF, 10000~km  (13 channels)\\
			&16D \cite{RademacherECOC2015}	&8~bit/16D-sym	& Experiments			& PM-BPSK &	BER 		&SNR: 2.5~dB  			& 55\% reach $\uparrow$, MCF, 14000~km (single channel)\\
			&24D \cite{millar2013SPPcom,millar2014OFC,Millar:14}	&12~bit/24D-sym	& Experiments			& PM-BPSK &	BER 		& SNR: 2--3~dB   			&15\% reach $\uparrow$,  25980~km (single channel)\\
			&Voronoi-based MD  \cite{MiraniJLT2021}	&24~bit/24D-sym	& AWGN/SSFM			& PM-$M$QAM &	MI/SER/BER		& SNR: 2--3~dB 			&  38\% reach $\uparrow$, at  BER of $2.26\times10^{-4}$\\
			\hline
			&\multicolumn{7}{l|}{SSFM: split-step Fourier method; MCF: multi-core fiber; DM: dispersion-managed; BICM-ID: bit interleaved coded modulation with iterative decoding; $\uparrow$: increase; $\downarrow$: decrease.} 
		\end{tabular}
		
}

%% file: pdf/NLIvsIter.tex
\hspace{-1em}
\subfigure{
\includegraphics{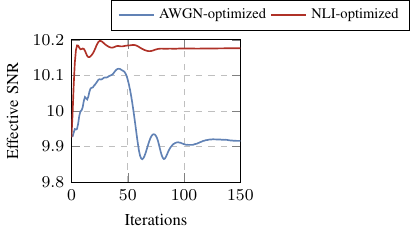}
}
\hspace{-8em}
\subfigure{
\includegraphics{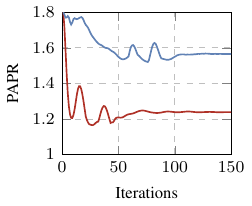}
}

\vspace{-1em}
\hspace{-1.7em}
\subfigure{
\includegraphics{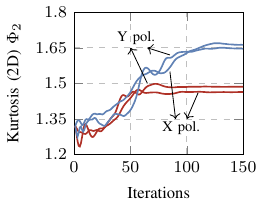}
}
\hspace{-0.8em}
\subfigure{
\includegraphics{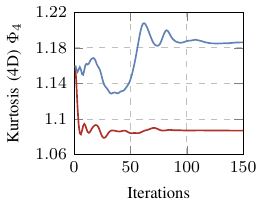}
}

\vspace{-1em}
\hspace{-1.65em}
\subfigure{
\includegraphics{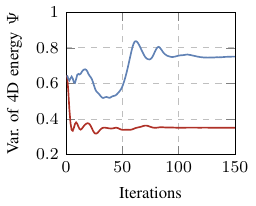}
}
\hspace{-0.7em}
\subfigure{
\includegraphics{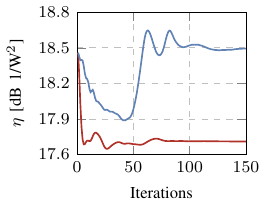}
}

%% file: Section_MultiSpan.tex
\section{AWGN-Optimized 4D Geometric Shaping for Optical Transmission}\label{sec:multi_span}

%In the following, we  describe the methodology used for the design of 4D modulation formats. %\footnote{\cb{Even the optimizations considered in this paper are limited to 4D, i.e. dual-polarization transmission, however, the optimization method is general to MD signals, e.g. joint time-slot constellation optimization or joint sub-carrier constellation shaping.}}
%The 4D modulation formats are optimized by taking into account both linear and nonlinear performance for a multi-span fiber system.
%The analysis focuses on GMI gains and the effect of shaping on the reach increase.

%\todo{11 channel}: 11 channel one format, 1 channel + 10 channel

%\todo{add the results without OS: 1) AWGN channel 2) in multi-span optical channel}

% \begin{table}[!tb]
% \footnotesize
%     \centering
% \caption{Simulation parameters for multi-span system.}\label{tab:para}
% \begin{tabular}{c|c}
% \hline\hline
% \multicolumn{2}{c}{\textbf{TX Parameters}} \\
% \hline
% Symbol rate & 45 Gbaud \\
% No. of WDM channels & 11 \\
% Channel spacing & 50 GHz \\
% Root-raised-cosine roll-off & 10\% \\
% \hline
% \multicolumn{2}{c}{\textbf{Fiber and Link Parameters}} \\
% \hline
% Attenuation coeff. ($\alpha$) & 0.21 dB/km \\
% Disp. parameter ($D$) & 16.9 ps/nm/km \\
% Nonlinear coeff. ($\gamma$) & 1.31 dB/km \\
% %\hline
% %\multicolumn{2}{c}{\textbf{Link Parameters}} \\
% %\hline
% Span length & 80 km \\
% EDFA noise figure & 5 dB \\
% \hline\hline
% \end{tabular}

% \end{table}

%In this paper, we xxxx   constraint of orthant-symmetry. xxxxx
In this section, we evaluate the performance of AWGN-optimized 4D constellations in the fibre-optic channel using SSFM simulations. To target a practical SD-FEC  with 20\%-25\% overhead, the optimizations were performed for the AWGN channel at an SNR for which $\text{GMI}\approx0.85m$~bit/4D-sym for six different SEs:  $m\in\{5,6,7,8,9,10\}$. %\footnote{The optimized 4D modulation formats can be found online at xx.} 
The recently proposed 4D modulation format with SE of  6~bit/4D-sym, i.e., 4D-64PRS \cite{BinChenJLT2019}, is chosen for $m=6$.
{In order to make the modulation more structured and reduce the optimization complexity as discussed in Sec.~\ref{sec:solutions}, the  OS  constraint is addied for $m=\{7, 8, 9, 10\}$.} 
For the transmission system and SD-FEC under consideration, the received SNR varies between 5~dB and 15~dB. 
Aiming at the  target of  $\text{GMI}\approx0.85m$~bit/4D-sym, 
 we optimize the six modulation formats  for a received SNR of 6, 8, 9.5, 10, 11 and 13 dB, respectively.
The geometrically-shaped 4D modulation formats with coordinates and labeling are  designed by solving Eq.~\eqref{GMI-optimal-OS} under the AWGN assumption and  using a gradient descent algorithm with the end-to-end autoencoder-learning approach in \cite{Kadir2019endtoend}. 

{The optimization process in detail for the geometrically-shaped 4D modulation formats  is as follows.
At the transmitter, binary vectors are first encoded as an ``one-hot" vector as input and  are mapped to constellation points via a neural network (NN).
We consider a simplified NN with no hidden layers, which only  consists of a fully-connected input layer and an output layer.  %without hidden layer 
%and the number of input and output neurons correspond to the constellation size and dimension.  
Normalization layer is performed to ensure the power constraint after the output layer.
The Gauss–Hermite quadrature is used for computing the GMI  as a cost function to remove stochastic effects and the OS constraint is added in the optimization   for reducing the complexity. 
For a special case of all zero biases,  the  weights directly correspond to the coordinates of the constellation points.
%Note that it has been shown that this approach is indeed sufficient to obtain state-of-the-art GMI-optimized constellations \cite{Kadir2019endtoend}.
These  weights  can be initialized with  random  variables or specific values.
However, initializing the neural network weights with Gray-labeled QAM is known to be good for
BICM systems and can accelerate the learning process compared to a random initialization. 
Therefore, we use the first orthant of PM-QAM and SP-QAM with Gray-labeling  for generating the initial constellations and  pre-training the initial weights\footnote{{Due to the limited number of constellation points,  the OS constraint is removed for optimizing 5~bit/4D-sym modulation format. The 4D 32-ary Voronoi constellation \cite{ForneyJSAC1989} with an optimized initial labeling  is used as the initial constellation.}}.
%For different target SEs,   various initial constellations are considered for pre-training the weight parameters, which is generated by using the orthant of PM-QAM and SP-QAM with Gray-labeling.\footnote{\changed{Due to the limited number of constellations points,  the OS constraint is removed for optimizing 5~bit/4D-sym modulation format. The 4D 32-ary Voronoi constellation \cite{ForneyJSAC1989} with an optimized initial labeling  is used as initial constellation.}}
The NN parameters %(i.e., weights and biases) 
are  optimized  using the Adam optimizer \cite{Kingma2014ICLR} with learning rate 0.01 to obtain the optimized coordinates of the constellation points.
}

\begin{figure*}[!tb]
\centering
{\includegraphics{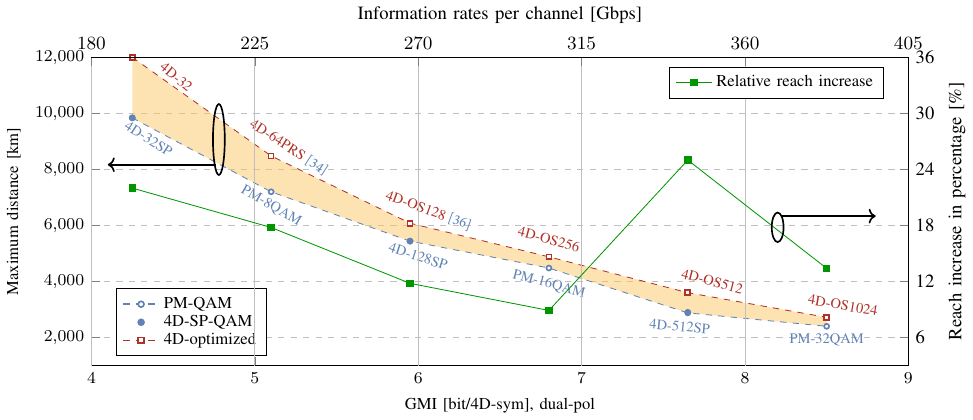}}
\vspace{-1em}
\caption{The maximum reach of various modulation formats for multi-span optical fiber transmission. 
The markers show the maximum transmission distance at $\text{GMI}=0.85m$ for  various 4D modulation formats. PM-QAM and 4D SP-QAM  are also shown as a reference.
} 
\label{fig:DvsSE}
\vspace{-0.5em}
\end{figure*}

\begin{figure}[!tp]
{\hspace{7em}  \footnotesize (a)   \hspace{12em}  \footnotesize (d) }

{\hspace{6.5em}  \footnotesize 4D-32%@10dB   
\hspace{10em}  4D-OS256}%@15dB}

\hspace{1.8em}\includegraphics[width=0.11\textwidth]{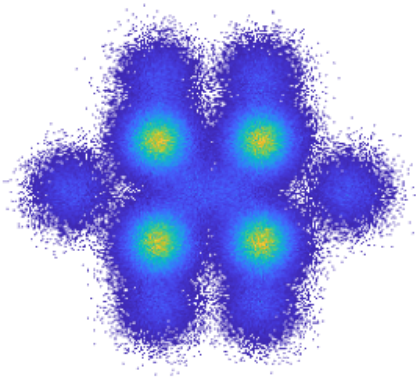}\hspace{-0.3em}
\includegraphics[width=0.11\textwidth]{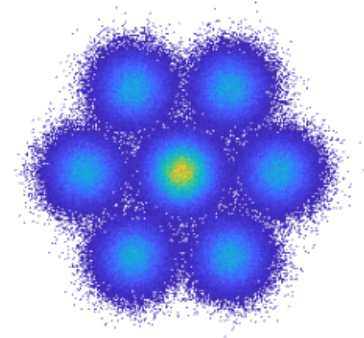} \hspace{0em}
\includegraphics[width=0.1\textwidth]{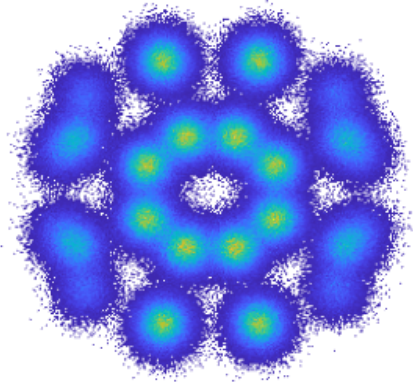}\hspace{0.1em}
\includegraphics[width=0.1\textwidth]{figure/256_apsk_os_constellation_noise_15dB_X.pdf}

{\hspace{7em}  \footnotesize (b)   \hspace{12em}  \footnotesize (e) }

{\hspace{6em}\footnotesize 4D-64PRS%@11.5dB  
\hspace{9em}  4D-OS512}%@16dB}

\hspace{2em}\includegraphics[width=0.1\textwidth]{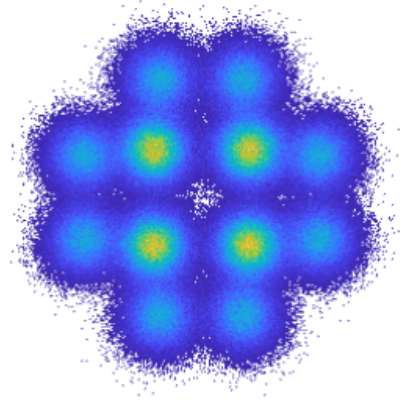}\hspace{0.1em}
\includegraphics[width=0.1\textwidth]{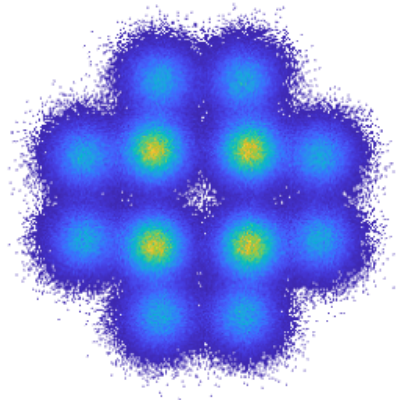} \hspace{0.2em}
\includegraphics[width=0.1\textwidth]{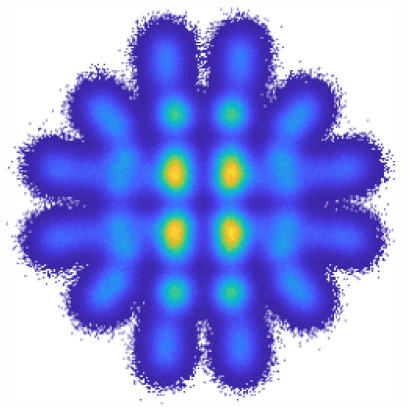}\hspace{0.1em}
\includegraphics[width=0.1\textwidth]{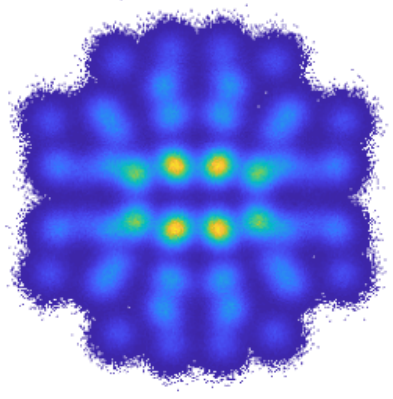}

{\hspace{7em}  \footnotesize (c)   \hspace{12em}  \footnotesize (f) }

{\hspace{6em}\footnotesize 4D-OS128%@13.5dB
\hspace{9em}  4D-OS1024}%@17dB}

\hspace{2em}\includegraphics[width=0.1\textwidth]{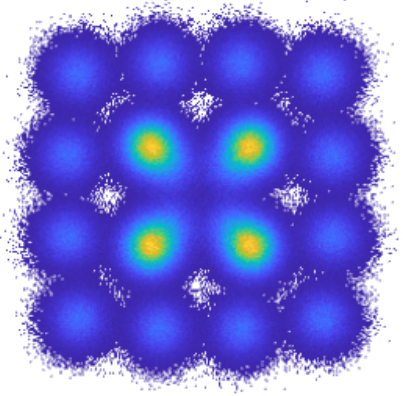}\hspace{0.1em}
\includegraphics[width=0.1\textwidth]{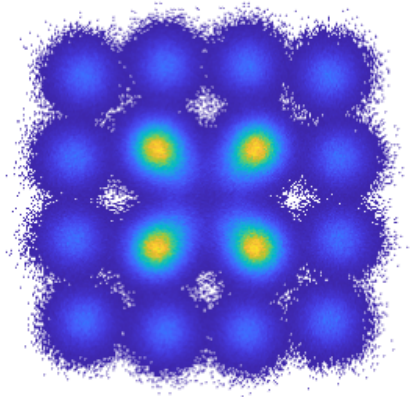} \hspace{0.2em}
\includegraphics[width=0.1\textwidth]{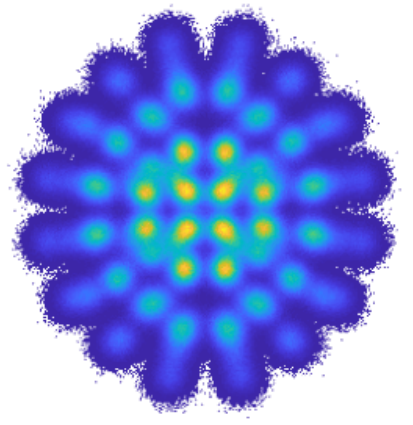}\hspace{0.1em}
\includegraphics[width=0.105\textwidth]{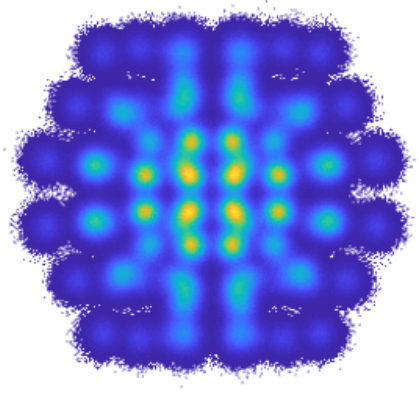}

\caption{Constellation diagrams of 4D modulation formats in $2\times$ 2D projection with SEs of 5--10~bit/4D-sym for a received    $\text{GMI}\approx0.95m$~bit/4D-sym as  (a) - (f).} %\todo{The optimized constellation coordinates and bit-to-symbol mappings are available for download at [xx].} %SE$=M\times0.85$. \todo{Update the results of QPSK and 16QAM, Or update with experiments?}
 %\todo{add the post-FEC BER of 6b2A8PSK and maybe also add the pre-BER in the figure.}}
%\todo{move the figure to Section 3 as a backtoback results.}}
\label{fig:modulations}
\vspace{-1em}
\end{figure} 

%\todo{The 4D-optimized modulations with $m=8, 9, 10$ are  obtained with the  constraint of OS.}
%\todo{Validation in optical channel:}
The considered system for evaluating the transmission performance of the modulation formats is a
dual-polarization long-haul WDM transmission system with 11 co-propagating channels over a multi-span of SSMF with attenuation parameter $\alpha=0.21$~dB/km, dispersion parameter $D =16.9$~ps$\cdot$nm$^{-1}\cdot$km$^{-1}$, and nonlinear coefficient $\gamma = 1.31$~W$^{-1}\cdot$km$^{-1}$. 
Each span consists of an 80~km SSMF through a split-step Fourier solution of the nonlinear Manakov equation with 800 steps per span and is followed by an erbium-doped fiber amplifier (EDFA) with noise figure of 5~dB.
Each WDM channel transmits a symbol rate of  45~Gbaud signal with channel spacing of 50~GHz and roll-off factor of 0.1 for root-raised-cosine  filter. 
At the receiver side, the DSP includes electronic chromatic dispersion compensation and matched filtering followed by ideal  phase rotation compensation. The performance of the center WDM channel is evaluated.
%The simulation parameters are given in Table \ref{tab:para} for the % multi-span
%optical fibre link under consideration.
%\todo{More details about the simulation setup, replace the table by text}

In Fig. \ref{fig:DvsSE}, the maximum transmission distance and the relative reach increase in percentage at $\text{GMI}=0.85m$~bit/4D-sym  of  twelve modulation formats are evaluated. 
%Fig. \ref{fig:DvsSE} shows the maximum transmission distance for various modulations formats at  $\text{GMI}=0.85m$ using a split-step Fourier method of the nonlinear Manakov equation for 11 WDM channels. %The parameters of this system are listed in Table \ref{tab:para}.
%, which comprises multiple standard single-mode fibre spans of 80~km, amplified at the end of each span by an EDFA with noise figure of 5~dB.
We observe that 4D-optimized  formats achieve approximately
320-2160~km (9\%-25\%) reach increase w.r.t  PM-QAM/4D-SP-QAM at the same information rates, which are highlighted by  the orange shaded region.
We note from Fig. \ref{fig:DvsSE} that 
larger reach increase in percentage  can be achieved w.r.t the QAM  modulations  without gray labeling.  %4D-SP-QAM. %, i.e. PM-8QAM and 4D SP-QAM. 
Especially comparing to  4D-SP32 and 4D-SP512,  the gains of  4D-optimized formats are more than 20\%, which is mainly due to the superior performance  of labeling.

Fig. \ref{fig:modulations}  shows the  optimized 4D modulation formats, i.e.,  4D-32, 4D-64PRS and 4D-OS$M$  (here, $M$ is taken to mean $M$-ary constellations),  for a received  GMI of $0.95\times m$~bit/4D-sym.\footnote{The optimized constellation coordinates and bit-to-symbol mappings are available at \href{https://github.com/TUe-ICTLab/Binary-Labeling-for-2D-and-4D-constellations}{https://github.com/TUe-ICTLab/Binary-Labeling-for-2D-and-4D-constellations}.}  
It shows the 4D symbol in each 2D projection,  whereas the symbol probability in 2D is proportional to the color brightness via yellow/blue colored heatmap. 
{It can be noticed that all the 4D modulations  have  symmetric shapes in each 2D, but they are not always X-pol/Y-pol symmetric. Here,  X-pol/Y-pol symmetry means that for all the 4D constellation points, if $[s_i^{1,2},s_i^{3,4}]\in \mathcal{S}$,  $[s_i^{3,4},s_i^{1,2}]$ must be  a valid 4D constellation point in $\mathcal{S}$.}
Interestingly, the  asymmetries  can especially be specially visible in  the higher cardinality configurations, i.e., 4D-OS512 and 4D-OS1024, but also for 4D-32.

%The shaded region highlight the range of reach increase over PM-QAM/4D-SP-QAM. 

%GMI in \ref{eq:GMI} of the constellations was then computed as a function of the transmission distance using the split-step Fourier method.

%% file: Section_SingleSpan.tex
\section{NLI Model-Aided 4D Geometric Shaping}
\label{sec:single_span}

As noted in the previous section, {all the modulation formats in Fig.~\ref{fig:DvsSE} are  designed for AWGN channel, with the exception of 4D-64PRS, which  uses the heuristic idea of constant-modulus constraint to improve the nonlinearity tolerance.}  
 NLI model is a key tool to analyze the performance of optical communication systems and enables the design of nonlinear-tolerant modulation formats.
 This is the objective of this section.

{The standard Gaussian noise (GN) model \cite{PoggioliniJLT2014, SerenaJLT2013,JohannissonJLT2013} ignores dependency
of nonlinear effects on the modulation format. 
Other more advanced models such as the enhanced Gaussian noise (EGN) model \cite{Carena:14,Dar:13} and NLI noise  model \cite{Dar:14} allow more accurate analysis of non-conventional  modulation formats by abandoning the gaussianity assumption of GN model.
However, these models were introduced in PM systems, so they are only valid for PM-2D modulation formats.% i.e. the symbols transmitted over two polarizations are statistically independent of each other and each polarization uses identical 2D modulation formats.
} 
For the nonlinear fiber channel, 4D NLI model  
 considering  modulation-dependent interference for all possible dual polarization four-dimensional (DP-4D) constellations have been  {proposed in \cite{GabrieleEntropy2020} %[HamiRabbani-JLT2021]
and evaluated in  \cite{GabrieleOFC2021,Zhiwei2022arxiv}.}
4D NLI model could provide a quick computation of the NLI power as a function of the input 4D modulation, which aims at ``shaping-out" the 4D modulation-dependent NLI term.%, as explained in \cite{Dar14_ISIT}. 

%The design should also aim at “shaping-out” the modulation-dependent NLIN term, as explained in [38]. To correctly shape-out the NLIN, a precise multi-dimensional model is required.

%For nonlinear fiber channel, NLI power models with considering modulation-dependent interference could provide a quick computation of the NLI power as a function of the input constellation, e.g., the enhanced Gaussian noise model \cite{Carena:14} for  PM-2D format and 4D NLI model \cite{GabrieleEntropy2020} for a general DP-4D format.

To evaluate the NLI,  the effective SNR, %(denoted by $\Gamma$), 
which represents the post-DSP SNR at the receiver,  is defined as %[xx]
\begin{align}\label{eq:SNR_eff}
\Gamma\triangleq\frac{P}{\sigma^{2}_{\text{ASE}}+\sigma^{2}_{\text{NLI}}},
\end{align}
where $P$, $\sigma^{2}_{\text{ASE}}$ and $\sigma^{2}_{\text{NLI}}$ represent the transmitted power, the variance of the amplified spontaneous emission
noise (ASE) and the NLI variance, respectively.

 For general 4D formats,   the NLI noise term $\sigma^2_{\text{NLI}}$ in Eq. \eqref{eq:SNR_eff} can be approximated as two nonidentical NLI powers over the two polarization channels as  {\cite [Eqs.~(42) and (43)]{GabrieleEntropy2020}}
\begin{align}\label{eq:noise_NLI}
 \sigma^2_{\text{NLI}}=\sigma^2_{\text{NLI,x}}+\sigma^2_{\text{NLI,y}}=\eta_{x}(\mathcal{P},\mathcal{S})P^3+\eta_{y}(\mathcal{P},\mathcal{S})P^3,
\end{align}
where $\eta_{x}(\cdot)$ and $\eta_{y}(\cdot)$ are functions of   a given system configuration $\mathcal{P}$ (fiber link parameters, launch power, etc.)  and the  4D modulation format $\mathcal{S}$, linked to the
contributions of the  modulation-dependent nonlinearities in the X- and Y-polarization, respectively.
The expression in Eq. \eqref{eq:noise_NLI} includes {X-pol and Y-pol interaction  effects} in the  nonlinear terms $\eta_x$ and $\eta_y$, which can be calculated via first-order regular perturbation in \cite[Eq. (42)-(43)]{GabrieleEntropy2020} and \cite[Eq. (1)]{GabrieleOFC2021}.\footnote{Note that in this paper, only signal-signal nonlinear interactions are considered along the fibre propagation in $\sigma^{2}_{\text{NLI}}$. For a more accurate NLI estimation,  
%especially for the system with nonlinearity compensation, 
signal-noise NLI interaction with  4D-modulation dependent contributions should be also considered, e.g., as done in \cite{Zhiwei2022arxiv}.}

Now that we have defined 4D modulation-dependent NLI and effective SNR, we turn our attention back to the GMI optimization in Eq. \eqref{eq:OP_GMI}.  %\eqref{GMI-optimal-OS}. 
Accordingly, to design a nonlinear-tolerant 4D modulation with the effective SNR under consideration, the new optimization problem for a given optical fiber channel parameters $\mathcal{P}$ can be reformulated as,
%SNR  depend on the modulation geometry
%{Extension of the MD Geometric Shaping with 4D model}
%Figure:  AWGN-learned modulation vs 4D model-learned  modulation \cite{GabrieleEntropy2020} %2) with iSABM
\begin{align}\label{eq:opt_4Dmodel}
	\{\mathcal{S}^*,\mathcal{B}^*\} & = \argmax_{\mathcal{S},\mathcal{B}%: E[||X||^2]\leq \sigma^2_x
	%: E[\|\bX\|^2]\leq \sigma_x^2
	} G\left(\mathcal{S},\mathcal{B},\Gamma_{\text{opt}}\right),
\end{align}
with 
\begin{align}\label{eq:SNR_opt}
\Gamma_{\text{opt}}\triangleq\frac{P_{\text{opt}}}{\sigma^{2}_{\text{ASE}}+\left(\eta_{x}(\mathcal{P},\mathcal{S})+\eta_{y}(\mathcal{P},\mathcal{S})\right)P^3_{\text{opt}}},
\end{align}
where $\Gamma_{\text{opt}}$ denotes the optimum effective SNR for a given system configuration  and  depends on the modulation format. The optimum value of the  launch power is obtained as  $P_{\text{opt}}=\sqrt[3]{\frac{\sigma^{2}_{\text{ASE}}}{2\left(\eta_{x}(\mathcal{P},\mathcal{S})+\eta_{y}(\mathcal{P},\mathcal{S})\right)}}$ by setting $\frac{\partial\Gamma}{\partial P}$  (from  Eq.~\eqref{eq:SNR_eff}) to zero.      %as a function of  $\mathcal{S}$ 
%for a given optical fiber channel parameters $\mathcal{P}$.
Thus, the optimum effective SNR in Eq. \eqref{eq:SNR_opt} can be rewriten as
\begin{align}\label{eq:SNR_opt2}
\Gamma_{\text{opt}}=\frac{2}{3\sqrt[3]{2}}\frac{1}{(\sigma^{2}_{\text{ASE}})^{2/3}\left(\eta_{x}(\mathcal{P},\mathcal{S})+\eta_{y}(\mathcal{P},\mathcal{S})\right)^{1/3}}
.
\end{align}
From the analysis above, we can see that  the GMI (see   Eq.~\eqref{eq:opt_4Dmodel},) is not only dependent on labeled
constellation $\{\mathcal{S},\mathcal{B}\}$, but also the  NLI noise via $\Gamma_{\text{opt}}$ in Eq.~\eqref{eq:SNR_opt2}. 
Note that for a fixed  parameter $\mathcal{P}$, the modulation-dependent NLI noise  is also a function of $\mathcal{S}$. Thus, the introduced NLI model-aided geometric shaping is actually only dependent on the shape of modulation formats and  without the need of SNR or power optimization, which is equivalent to add a constraint and reduce the dimensionality of the optimization space.

% In this paper, we solve Eq. \eqref{eq:opt_4Dmodel} with \cb{the OS constraint (mentioned in Sec. \ref{subsec:solutions})} using  end-to-end learning
% following \cite{Kadir2019endtoend} by maximizing GMI.
% The 4D NLI model can be considered as a surrogate channel in the end-to-end learning structure, thus,  gradient descent need to be applied to both constellation coordinates and  effective SNR.
% Similar 4D NLI model-aided 4D modulation optimization  has been done in [GabrieleOFC2022], but target on symbol-wise coded modulation system by maximizing mutual information (MI) for a multi-span transmission scenario.

%\todo{Describe the simulation system (single-span) we consider for 4D model} 
We consider a dual-polarized, single
channel waveform  over a single span of SSMF. 
The same fiber parameters (including dispersion, attenuation and nonlinearity) as in Sec. \ref{sec:multi_span} are used for both 4D NLI model and SSFM simulation.
We solve Eq.~\eqref{eq:opt_4Dmodel}  {by adding the OS constraint% as Eq. \eqref{GMI-optimal-OS}  in Sec. \ref{subsec:solutions}
} and also using the end-to-end AE-learning method described before. %\footnote{\todo{In order to make the modulation more structured and reduce the optimization complexity,  constraint of OS  is added in the optimization as Eq. \eqref{GMI-optimal-OS}.}}
 The 4D NLI model can be considered as a surrogate channel in the end-to-end learning structure, thus,  gradient descent needs to be applied to constellation coordinates and  optimum effective SNR both for  maximizing GMI. 
Similar 4D NLI model-aided 4D modulation optimization  has been also done in \cite{GabrieleOFC2022}, however lifting any geometrical constraint and targeting only symbol-wise coded modulation systems by maximizing MI for a multi-span transmission scenario.

\begin{figure}[!tb]
\centering
{\includegraphics{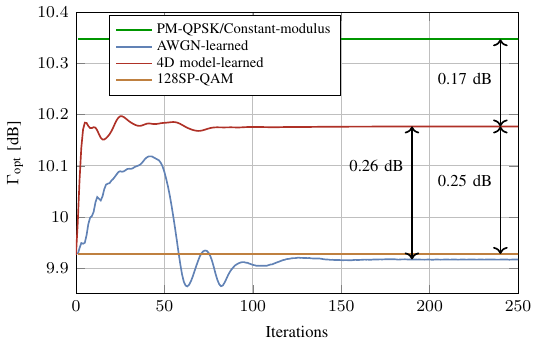}}
\vspace{-1.5em}
\caption{The   optimum effective SNR $\Gamma_{\text{opt}}$ versus the number of optimization iteration  at 234~km  of geometrically shaped 4D 128-ary modulation format optimized based on AWGN channel and 4D NLI model.  %Results of modulation optimization for $N=4$ and $M=128$ The optimization is done separately for AWGN channel and  4D NLI model. The 4D NLI model  considers a single-span  optical transmission system with single channel. (a) $\text{SNR}_{\text{opt}}$  vs. the number of iterations at 234~km; (b) GMI vs. the number of iterations.
} 
\label{fig:4Dmodel_opt1}
\vspace{-1em}
\end{figure}

\begin{figure}[!tb]
\centering
{\includegraphics{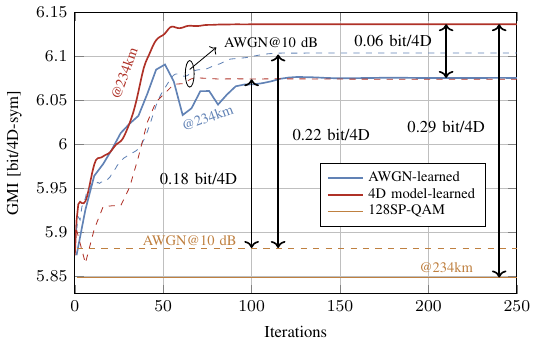}}
\vspace{-1.5em}
\caption{GMI versus the number of optimization iteration  at 234~km   of geometrically shaped 4D 128-ary modulation format optimized based on AWGN channel and 4D NLI model.  %Results of modulation optimization for $N=4$ and $M=128$ The optimization is done separately for AWGN channel and  4D NLI model. The 4D NLI model  considers a single-span  optical transmission system with single channel. (a) $\text{SNR}_{\text{opt}}$  vs. the number of iterations at 234~km; (b) GMI vs. the number of iterations.
} 
\label{fig:4Dmodel_opt2}
\vspace{-1em}
\end{figure}

%\todo{Optimization comparison}
In Fig.  \ref{fig:4Dmodel_opt1} and Fig.  \ref{fig:4Dmodel_opt2},  the optimum effective SNR and GMI trace as a function of the optimization steps are shown for the optimization procedure.
The 4D modulation  formats with a SE of 7~bit/4D-sym  are optimized with OS constraint via end-to-end learning following \cite{Kadir2019endtoend} by maximizing GMI.
The  simulations  are  implemented  by  solving two  optimization  problems with  similar effective SNRs around 10~dB: one is for the  AWGN channel with SNR=10~dB (AWGN-learned) and the other is for 4D NLI-model \cite{GabrieleEntropy2020} with a single-channel, 234~km single-span transmission system (4D model-learned).   
%We initialized the optimization with .
PM-QPSK as a format with a good nonlinearity tolerance and 4D-128SP-QAM \cite{ErikssonOE13} with 7~bit/4D-sym  are shown as references.

Fig.~\ref{fig:4Dmodel_opt1} shows that the 4D model-learned modulation can tolerate higher nonlinearities.  This can be seen from the  achieved  0.25~dB %(0.42~dB)
gain with respect to 128SP-QAM and AWGN-learned 4D modulation in  terms  of  $\text{SNR}_{\text{opt}}$  at 234~km. 
In contrast, PM-QPSK or 4D constant-modulus modulation formats show that   a  potential effective SNR  gain of  0.42~dB for nonlinearity tolerance could be provided. The 4D model-learned modulation   achieves more than half of this gain, which is translated  into a  GMI increase as explained below.

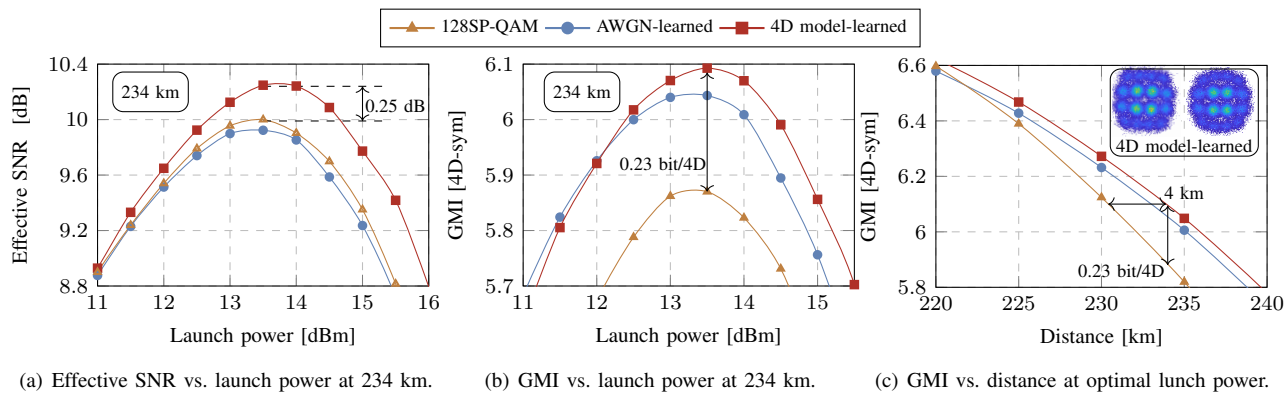
\begin{figure*}[!tb]
\vspace{-0.5em}
\centering
\input{./pdf/comparison_234km.tex}
\vspace{-0.3em}
\caption{Simulation results of single-span optical fiber transmission with single channel for three modulation formats: 128SP-QAM, AWGN-learned 4D modulation and 4D model-learned modulation. %\todo{Add the constellation as insets in (c).}
}% (a) Launch Power vs. SNR at 234~km; (b) Launch Power vs. GMI at 234~km; (c) GMI vs. distance. } 
\label{fig:single_channel}
\vspace{-1em}
\end{figure*}

Fig.~\ref{fig:4Dmodel_opt2}  shows that in an AWGN channel  with an SNR of 10~dB (dashed curves),  both AWGN-learned modulation format and 4D model-learned modulation format outperform   128SP-QAM with a gain of 0.22~bit/4D and  0.18~bit/4D, respectively.
%provide the gain around 0.22~bit/4D (in term of GMI)  with respect to 4D-128SP-QAM, while the gain is around 0.18~bit/4D for 4D-learned modulation format. 
However, by considering  an optical fiber channel with length of 234~km (solid curves),   the gain of the 4D model-learned modulation  is increased to 0.29~bit/4D compared to
128SP-16QAM, which is higher than that of the AWGN-learned format. 
These benefits (approximately 0.11~bit/4D GMI gain) come from the improvement of $\text{SNR}_{\text{opt}}$ shown in Fig.~\ref{fig:4Dmodel_opt1} due to its excellent nonlinear-tolerant property.
The relative additional shaping gain in
terms of GMI for the nonlinear channel with respect to the gain in the linear channel is $(0.29-0.18)/0.18=38\%$. 
%Note that both  the AWGN-learned  format and 128SP-QAM lead to an observable effective SNR degradation  in nonliear optical channel with respect to AWGN channel
%, and approximately 0.11~bit/4D gain comes from the $\text{SNR}_{\text{opt}}$ improvement.
%the 4D model-learned modulation can  provide the  gain  of  approximately 0.286~bit/4D and 0.074~bit/4D  at 234~km  with  respect  to  4D-128SP-QAM and the AWGN-learned modulation, respectively. In contrast, the gains in AWGN channel with fixed SNR of 10~dB is much smaller.
It well-matched to the fact that 4D model-learned modulations  lead to a good trade-off between linear and nonlinear shaping gain  by  increasing the linear shaping gain and  maintaining a fair level of nonlinearity tolerance.

In Fig.~\ref{fig:single_channel},  we show the effective SNR and GMI performance of three modulation formats  through a split-step Fourier solution of the nonlinear Manakov equation. 
We can observe that the 4D model-learned  modulation outperforms both 128SP-QAM and AWGN-learned 4D modulation   in term of effective SNR and GMI, which is close to the observed gain in 4D NLI model prediction in Fig.~\ref{fig:4Dmodel_opt1} and Fig.~\ref{fig:4Dmodel_opt2}.
Normally, a GS modulation format will lead to a larger SNR penalty. 
However, the 4D model-learned modulation format provides a larger effective SNR (0.25~dB gain) compared to 128SP-16QAM at the optimal launch power in Fig.~\ref{fig:single_channel} (a), which is consistent with the analysis in Sec. II-C.
In addition, both  the AWGN-learned  format and 128SP-QAM lead to a larger effective SNR degradation   than 4D model-learned modulation format in nonlinear region (high launch power).
Therefore, the shaping gain in the linear region can be increased in nonlinear region and translates into a larger reach increase.
 {In Fig.~\ref{fig:single_channel} (b), we can also observe that the GMI gain of AWGN-learned modulation over 4D model-learned modulation in linear region is vanished  in nonlinear region.}

Fig.~\ref{fig:single_channel} (c) shows GMI as a function of transmission distance for
three different modulation formats using the optimal launch power at each distance. 
{The received symbols in 2D projection (X-pol/Y-pol) of the 4D model-learned modulation is plotted in the inset  of Fig.~\ref{fig:single_channel} (c).}
We can observe that the proposed 4D model-learned modulation format leads to a 4~km (2\%) increase in reach relative to the 128SP-16QAM modulation format at GMI of 6.1~bit/4D-sym.
In addition, 4D model-learned format performs better than AWGN-learned  format  with a gain of about 1~km in all the distances between 220~km and 240~km.

%The red and blue circles represent the symbols transmitted in X and Y polarization respectively.

%a gain of 1~km in transmission distance between 4D model learned format and AWGN-learned modulation format is achieved.

%4D model learned  modulation format perform better than AWGN-learned  modulation format in all the distance.

 %Fig.~\ref{fig:single_channel} (c) shows that the reach increase of xx at optimal launch power is  xx~km larger than xx. A gain of xx\% in transmission distance between xx and xx  is achieved at the GMI of 6~bit/4D-sym.

% \todo{a Fully loaded DWDM Systems}
% \begin{itemize}
%     \item 1 vs 11 vs 41 or 81 WDM channels
% \end{itemize}

%% file: pdf/comparison_234km.tex
\hspace{-1em}
\subfigure[Effective SNR  vs. launch power at 234~km.]{
 {\includegraphics{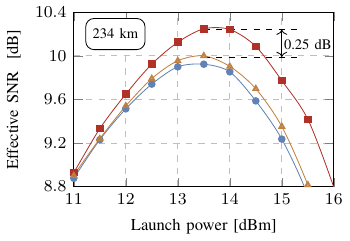}}
}
\hspace{-4em}
\subfigure[GMI vs. launch power at 234~km.]{
{\includegraphics{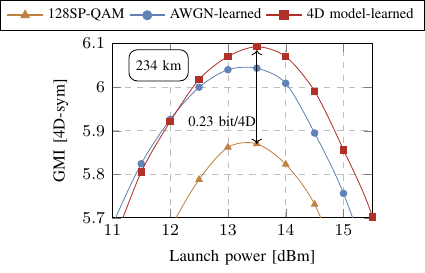}}
}
\hspace{-4em}
\subfigure[GMI vs.  distance at optimal lunch power.]{
{\includegraphics{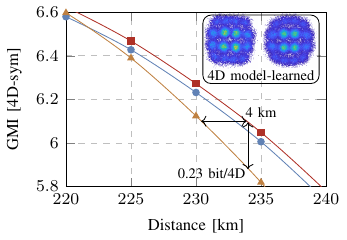}}
}